\documentclass[12pt,a4paper]{article}
\usepackage[utf8]{inputenc}
\usepackage{epsfig}
\usepackage{mathtools}
\usepackage{amsfonts}
\usepackage{amssymb}
\usepackage{amsmath}
\usepackage{cancel}
\usepackage{color}
\usepackage{slashed}
\usepackage{cite}
\usepackage{caption}
\usepackage{subcaption}
\usepackage{comment}
\usepackage{marginnote}

\usepackage{tikz}
\usetikzlibrary{shapes}
\usetikzlibrary{trees}
\usetikzlibrary{calc}
\usetikzlibrary{decorations.pathmorphing}
\usetikzlibrary{decorations.markings}

\allowdisplaybreaks

\def\FF              {\mathcal{F}}

\def\bP{{\bf P}}

\def\s{\sigma}

\def\bp{{\bf p}}
\def\btp{{\bf\tilde p}}
\def\btP{{\bf\tilde P}}
\def\bP{{\bf P}}
\def\tbP{\tilde{\bP}}
\def\OO{\mathcal{O}}
\def\Z{\mathcal{Z}}
\def\P{\mathcal{P}}

\newcommand{\barH}{{\bar H}}

\def\B{\mathcal{B}}

\def\j{\bullet}

\begin{document}

\begin{center}

\vspace{1cm}

{ \Large\bf ABJM quantum spectral curve at twist 1: \\
	algorithmic perturbative solution} \vspace{1cm}

{\large R.N. Lee$^{1}$ and  A.I. Onishchenko$^{2,3,4}$}\vspace{0.5cm}

{\it $^1$ Budker Institute of Nuclear Physics, Novosibirsk,
	Russia,\\
	$^2$Bogoliubov Laboratory of Theoretical Physics, Joint
	Institute for Nuclear Research, Dubna, Russia, \\
	$^3$Moscow Institute of Physics and Technology (State University), Dolgoprudny, Russia\\
	$^4$Skobeltsyn Institute of Nuclear Physics, Moscow State University, Moscow, Russia}\vspace{1cm}

\abstract{
 We present an algorithmic perturbative solution of ABJM quantum spectral curve at twist 1 in $sl(2)$ sector for arbitrary spin values, which can be applied to, in principle, arbitrary order of perturbation theory. We determined the class of functions --- products of rational functions in spectral parameter with sums of Baxter polynomials and Hurwitz functions --- closed under elementary operations, such as shifts and partial fractions, as well as differentiation. It turns out, that this class of functions is also sufficient for finding solutions of inhomogeneous Baxter equations involved. For the latter purpose we present recursive construction of the dictionary for the solutions of Baxter equations for given inhomogeneous parts. As an application of the proposed method we present the computation of anomalous dimensions of twist 1 operators at six loop order. There is still a room for improvements of the proposed algorithm related to the simplifications of arising sums. The advanced techniques for their reduction to the basis of generalized harmonic sums will be the subject of subsequent paper. We expect this method to be generalizable to higher twists as well as to other theories, such as $\mathcal{N}=4$ SYM.}

\end{center}

\begin{center}
	ABJM, Baxter equations
\end{center}

\newpage

\tableofcontents{}\vspace{0.5cm}

\renewcommand{\theequation}{\thesection.\arabic{equation}}

\section{Introduction}

The study of integrability in quantum field theories with extended supersymmetry in spacetime dimensions greater then two is quite young subject. Nevertheless, we already have a plenty of results, for a review and introduction see Ref. \cite{IntegrabilityReview,IntegrabilityPrimer,IntegrabilityDeformations,IntegrabilityDefects,IntroductionQSC,IntegrabilityStructureConstants,FishnetCFTreview}. Initially, this study was started with the so called ``experimental'' tests of  AdS/CFT duality \cite{tHooftDuality,MaldacenaAdSCFT,GKP,WittenAdSHolography} and soon it was realized that many techniques from the world of two-dimensional integrable systems, such as sigma-model and spin-chain S-matrices \cite{StaudacherSMatrix,BetheAnsatzQuantumStrings,BeisertDynamicSmatrix,BeisertAnalyticBetheAnsatz,TranscedentalityCrossing,JanikWorldsheetSmatrix,ArutyunovFrolovSmatrix,ZamolodchikovFaddevAlgebraAdS,N6Smatrix}, Asymptotic Bethe Ansatz (ABA) \cite{MinahanZarembo,N4SuperSpinChain,LongRangeBetheAnsatz,TranscedentalityCrossing,MinahanZaremboChernSimons,SpinChainsN6ChernSimons,AllloopAdS4}, Thermodynamic Bethe Ansatz (TBA) \cite{TBAN4,TBAN4proposal,TBAexcitedstates,TBAMirrorModel} as well as $Y$ and $T$-systems \cite{YsystemAdS5,TBAfromYsystem,WronskianSolution,SolvingYsystem,TBAYsystemAdS4,GromovYsystemAdS4,DiscontinuityRelationsAdS4} can be also applicable for the computation of anomalous dimensions of various operators in these theories. The most well understood theories at a moment are $\mathcal{N}=4$ SYM in four and $\mathcal{N}=6$ super Chern-Simons theory (ABJM model) in three dimensions \cite{ABJM1}.  The integrability based methods were also used in the study of quark-antiquark potential \cite{potentialTBA,IntegrableWilsonLoops,cuspQSC,potentialQSC}, expectation values of polygonal Wilson loops at strong coupling and beyond \cite{BubbleAnsatz,YsystemScatteringAmplitudes,OPEpolygonalWilsonLines,SmatrixFiniteCoupling,OPEHelicityAmplitudes,OPEHelicityAmplitudes2,POPEresummation,AsymptoticBetheAnsatzGKPvacuum,StrongCouplingPOPE,ABJMflux-tube}, eigenvalues of BFKL kernel \cite{adjointBFKL,GromovBFKL1,GromovBFKL2,BFKLnonzeroConformalSpin}, structure constants \cite{StructureConstantsPentagons,StructureConstantsWrappingOrder,ClusteringThreePointFunctions,StructureConstantsLightRayOperators,QSC_StructureConstants}, correlation functions \cite{StructureConstantsPentagons,CorrelationFunctionsIntegrability1,CorrelationFunctionsIntegrability2,CorrelationFunctionsIntegrability3,CorrelationFunctionsIntegrability4,CorrelationFunctionsIntegrability5,CorrelationFunctionsIntegrability6,CorrelationFunctionsIntegrability7,CorrelationFunctionsIntegrability8,CorrelationFunctionsIntegrability9,CorrelationFunctionsIntegrability10}, one-point functions of operators in the defect conformal field theory \cite{defectCFT1,defectCFT2,defectCFT3} and observables at finite temperature such as Hagedorn temperature of $\mathcal{N}=4$ SYM \cite{HagedornTemperatureIntegrability1,HagedornTemperatureIntegrability2}.

Eventually, a detailed study  of TBA equations for super spin chains corresponding to  $\mathcal{N}=4$ SYM and ABJM models has led to their simplified alternative formulations in terms of Quantum Spectral Curve (QSC), a set of algebraic relations for Baxter type $Q$-functions together with analyticity and Riemann-Hilbert monodromy conditions for the latter \cite{N4SYMQSC1,N4SYMQSC2,twistedN4SYMQSC,N4SYMQSC3,N4SYMQSC4,ABJMQSC,ABJMQSCdetailed,QSCetadeformed}. Within the quantum spectral curve formulation one can relatively easy obtain numerical solution for any coupling and state \cite{QSCnumericsN4SYM1,QSCnumericsN4SYM2,QSCnumericsABJM}. Also, QSC formulation allowed to construct iterative perturbative solutions for these theories at weak coupling up to, in principle, arbitrary loop order \cite{VolinPerturbativeSolution,ABJMQSC12loops,N4SYMQSC4}. The mentioned solutions are however limited to the situation when the states quantum numbers are given explicitly by some integer numbers.  Recently, in Ref. \cite{ABJM_QSC_Mellin,uspaceTMF} we started developing techniques for the solution of QSC equations  treating state quantum numbers as symbols. The first technique based on Mellin space transform \cite{ABJM_QSC_Mellin} turned out to be quite complex to go for all-loop generalization. On the other hand, in Ref. \cite{uspaceTMF} we suggested, that there should be relatively easy way to obtain a perturbative solution for the spectrum of twist 1 operators in $sl(2)$ sector for ABJM model working directly in spectral parameter space. The goal of this paper is to  present the algorithm for perturbative solution of ABJM quantum spectral curve at twist 1 in $sl(2)$ sector to any loop order. The latter is based on the existence of a class of functions - products of rational functions in spectral parameter with sums of Baxter polynomials and Hurwitz functions, which is closed under elementary operations, such as shifts and partial fractions,  as well as differentiation. The introduced class of function is sufficient for finding solutions of involved inhomogeneous Baxter equations using recursive construction of the dictionary for the solutions of Baxter equations for given inhomogeneous parts. As an application of the proposed method we present computations of anomalous dimensions of twist 1 operators at six loop order. There is still a room for improvements of the proposed algorithm related to the simplifications of the arising sums and we plan to present advanced techniques for their reduction to the basis of generalized harmonic sums in one of our subsequent papers. The presented approach  has the potential for generalizations to higher twists of operators, as well as to other theories such as $\mathcal{N}=4$ SYM and twisted $\mathcal{N}=4$ SYM and ABJM models.

This paper is organized as follows.  In the next section we
give necessary details on ABJM quantum spectral curve equations putting emphasis on $\bP\nu$-system. Section \ref{Riemann-Hilbert-solution} contains all the details about our solution of Riemann-Hilbert problem for $\bP\nu$-system, used for calculation of anomalous dimensions of twist 1 operators. 
Next, section \ref{AnomalousDimensions} contains the results for anomalous dimensions up to six loop order and their discussion. Finally, in section \ref{Conclusion} we come with our conclusion.  Appendices and {\it Mathematica} notebooks contain some details of our calculation.

\section{ABJM quantum spectral curve}\label{ABJM-QSC}

As it was already mentioned in introduction, the ABJM model is the second most popular playground for testing AdS/CFT correspondence. It is 
a three-dimensional $\mathcal{N}=6$ Chern-Simons theory based on the product $U(N)\times \hat{U}(N)$ of two gauge groups with opposite Chern-Simons levels $\pm k$. In the planar limit, where $N, k\to\infty$ so that the 't Hooft coupling $\lambda = \frac{k}{N}$ kept fixed, this theory has a dual description in terms of IIA superstring theory on $AdS_4\times CP^3$. The field content of ABJM model consists of two gauge fields $A_{\mu}$ and $\hat{A}_{\mu}$, four complex scalars $Y^A$ and four Weyl spinors $\psi_A$ with matter fields transforming in the bi-fundamental representation of the gauge group. The global symmetries of ABJM theory with Chern-Simons level $k > 2$ are given by orthosymplectic supergroup $\text{OSp} (6|4)$ \cite{ABJM1,ABJM2} and the ``baryonic'' $U(1)_b$ \cite{ABJM2}. The bosonic subgroups of $\text{OSp} (6|4)$ supergroup are related to isometries of superstring background $AdS_4\times CP^3$.   

In the present paper we will be interested in the calculation of anomalous dimensions of twist 1 gauge-invariant operators from $sl (2)$ sector for arbitrary spin values $S$. The latter are given by single-trace operators of the form\cite{KloseABJMreview}:
\begin{eqnarray}
\text{tr} \left[ D_{+}^S (Y^1 Y_4^\dagger)^L\right] . \label{sl2op}
\end{eqnarray}
where twist 1 corresponds to $L=1$. The expressions for anomalous dimensions can be conveniently obtained by solving the corresponding spectral problem for $\text{OSp} (6|4)$ spin chain. The most advanced framework for that at the moment is offered by quantum spectral curve (QSC) method. The latter is an alternative reformulation of Thermodynamic Bethe Ansatz (TBA) as a finite set of functional equations: $Q$-system. The most important advance is provided by the considerable simplification of the spectral problem calculations. In the case of ABJM model QSC formulation was introduced in Ref. \cite{ABJMQSC,ABJMQSCdetailed}, see also Ref. \cite{ABJMQSC12loops}. To perform actual calculations of anomalous dimensions we will use monodromy conditions for the part of ABJM Q-system known as $\bP\nu$-system \cite{ABJMQSC,ABJMQSCdetailed}. The latter consists of six functions ${\bf P}_A, A=1,\ldots ,6$ and eight ($4+4$) functions  $\nu_a, \nu^b, a,b=1,\ldots 4$ satisfying nonlinear matrix Riemann-Hilbert problem \cite{ABJMQSC,ABJMQSCdetailed}:
\begin{align}
\tbP_{ab} - \bP_{ab} &= \nu_a\tilde{\nu}_b - \nu_b\tilde{\nu}_a\, ,\quad &\tbP^{ab} - \bP^{ab} &= -\nu^a\tilde{\nu}^b + \nu^b\tilde{\nu}^a\, , \\
\tilde{\nu}_a &= -\bP_{ab}\nu^b\, ,\quad &\tilde{\nu}^a &= -\bP^{ab}\nu_b\, ,
\end{align} 
where
\begin{eqnarray}
\bP_{ab} = \left( \begin{array}{cccc} 0 & - \bP_1 & - \bP_2 & - \bP_5 \\  \bP_1 & 0 & - \bP_6 & - \bP_3 \\ \bP_2 & \bP_6 & 0 & - \bP_4 \\ \bP_5 & \bP_3 & \bP_4 & 0 \end{array} \right) , \;\;\;\; \bP^{ab} = \left( \begin{array}{cccc} 0 & \bP_4 & - \bP_3 & \bP_6 \\  -\bP_4 & 0 & \bP_5 & - \bP_2 \\ \bP_3 & -\bP_5 & 0 & \bP_1 \\ -\bP_6 & \bP_2 & -\bP_1 & 0 \end{array} \right) .
\end{eqnarray}
Here and in the following  $\tilde{f}$ will denote a function $f$ analytically continued around one of the branch points on the real axis.
In addition, the $\bP$ and $\nu$ - functions should satisfy extra constraints
\begin{eqnarray}
\bP_5 \bP_6 &=& 1 + \bP_2 \bP_3 - \bP_1 \bP_4, \label{Pconstr}\\
\nu^a\nu_a &=& 0 , \label{nuconstr}
\end{eqnarray}
Both ${\bf P}_A$ and $\nu_a, \nu^a$ are functions of spectral parameter $u$. The ${\bf P}_A$ functions have a single cut on the defining Riemann sheet running from  $-2 h$ to $+2 h$ ($h$ is effective ABJM QSC coupling constant\footnote{In contrast to $\mathcal{N}=4$ SYM, ABJM QSC coupling constant $h$ is a nontrivial function of 't Hooft coupling constant $\lambda$, Ref. \cite{SpinChainsN6ChernSimons,StringDualN6ChernSimons}. There is a conjecture for the exact form of $h(\lambda)$, Ref. \cite{hconjecture1,hconjecture2}, made by a comparison with the localization results.}), while $\nu_a, \nu^a$ functions have an infinite set of branch cuts located at intervals $(-2h,+2h) + i n, n\in \mathbb{Z}$ and satisfy  simple  quasi-periodicity relations
\begin{equation}
\tilde{\nu}_a (u) = e^{i\mathcal{P}}\nu_a (u+i)\, ,\quad 
\tilde{\nu}^a (u) = e^{-i\mathcal{P}}\nu^a (u+i)\, ,
\end{equation}
where $\mathcal{P}$ is a state dependent phase fixed from self-consistency of QSC equations \cite{ABJMQSCdetailed}. To get QSC description of states in $sl(2)$ sector (\ref{sl2op}) it is sufficient to consider ${\bf P}\nu$-system reduced to symmetric, parity invariant states. The reduced ${\bf P}\nu$-system is identified by constraints $\bP_5 = \bP_6 = \bP_0$, $\nu^a = \chi^{ab}\nu_b$ and is written as \cite{ABJMQSC,ABJMQSCdetailed,ABJMQSC12loops}:
\begin{eqnarray}
{ \tilde \nu}_a &=& - \bP_{ab} \, \chi^{bc} \nu_c  \label{nuPmonodromy} ,\\
{ \widetilde {\bP} }_{ab} - { \bf P }_{ab} &=& \nu_a { \tilde \nu }_b - \nu_b { \tilde \nu }_a , \label{Pdiscont} 
\end{eqnarray}  
where
\begin{eqnarray}
\chi_{ab} = \left( \begin{array}{cccc} 0 & 0 & 0& - 1 \\  0 & 0 & 1 & 0 \\ 0 & -1 & 0 & 0 \\ 1 &  0& 0 & 0 \end{array} \right) .
\end{eqnarray}
and $\nu_a$  satisfy now periodic/anti-periodic constraints ($\sigma = \pm 1$)
\begin{eqnarray}
{ \tilde \nu }_a = e^{i\mathcal{P}} \nu_a^{[2]} = \sigma \, \nu_a^{[2]}\, , \label{nuanalytcont}
\end{eqnarray}
where $f^{\left[n\right]}\left(u\right)  =f\left(u+in/2\right)$ and the constraint for $\bP$ functions takes the form
\begin{eqnarray}
( \bP_0 )^2 &=& 1 - \bP_1 \bP_4 + \bP_2 \bP_3 . \label{P0}
\end{eqnarray}
In addition to the above analytical properties of $\bP$ and $\nu$ functions it is required \cite{ABJMQSC12loops,ABJMQSCdetailed} that they are free of poles and stay bounded at branch points. The quantum numbers of spin chain states under consideration, that is twist $L$, spin $S$ and conformal dimension $\Delta$ are encoded in the behavior of $\bP$, $\nu$ functions at large values of spectral parameter $u$ \cite{ABJMQSC,ABJMQSCdetailed,ABJMQSC12loops}:  
\begin{eqnarray}
\bP_a &\simeq& ( A_1 u^{-L}, \,A_2 u^{-L-1}, \,A_3 u^{+L+1}, \,A_4 u^{+L}, \, A_0 u^0) , \nonumber\\
A_1 A_4 &=&-\frac{(\Delta -L+S) (\Delta -L-S+1) (\Delta +L-S+1) (\Delta +L+S)}{L^2 (2 L+1)} \nonumber\\
A_2 A_3 &=&-\frac{(\Delta -L+S-1) (\Delta -L-S) (\Delta +L-S+2) (\Delta +L+S+1)}{(L+1)^2 (2 L+1)},\label{Pasympt}
\end{eqnarray}
and  
\begin{eqnarray}
\nu_a \sim \left( u^{\Delta - L} ,u^{\Delta+1}, u^{\Delta} , u^{\Delta+L+1}  \right)\, , 
\end{eqnarray}
which serve as boundary conditions for the Riemann-Hilbert problem under study. 
The anomalous dimension $\gamma$, which is our main interest here, is given by $\gamma = \triangle - L - S$.

\section{Solution of Riemann-Hilbert problem for $\bP\nu$-system}\label{Riemann-Hilbert-solution}

To solve the Riemann-Hilbert problem for fundamental $\bP\nu$-system it is convenient to add to original equations \eqref{nuPmonodromy} - \eqref{Pdiscont} their algebraic consequences \cite{ABJMQSC12loops}. First, from Eq. \eqref{nuPmonodromy} using Eq. \eqref{P0} we obtain
\begin{equation}
\nu_a = -\bP_{ab}\chi^{bc}\tilde{\nu}_c \label{nuPmonodromy2}
\end{equation}
Next, dividing by $\bP_{12}^{[2]}$ equation \eqref{nuPmonodromy} and subtracting from it the equation \eqref{nuPmonodromy2} divided by $\bP_{12}$ we get 
\begin{equation}
\frac{\nu_a^{[3]}}{\bP_{12}^{[1]}} -  \frac{\nu_a^{[-1]}}{\bP_{12}^{[-1]}} + \sigma \left(
\frac{\bP_{ab}^{[1]}}{\bP_{12}^{[1]}} - \frac{\bP_{ab}^{[-1]}}{\bP_{12}^{[-1]}}
\right)\chi^{bc}\nu_c^{[1]} = 0
\end{equation}
We will also need the equations following from the sum of equations \eqref{nuPmonodromy} and \eqref{nuPmonodromy2}:
\begin{align}
\left(
\nu_1 + \s \nu_1^{[2]}
\right) \left(
\bp_0 - (h x)^L
\right) &= \bp_2 \left(
\nu_2 + \s \nu_2^{[2]}
\right) - \bp_1 \left(
\nu_3 + \s \nu_3^{[2]}
\right)\, , 
\\ 
\left(
\nu_2 + \s \nu_2^{[2]}
\right)\left(
\bp_0 + (h x)^L
\right) &= \bp_3 \left(
\nu_1 + \s \nu_1^{[2]}
\right) + \bp_1 \left(
\nu_4 + \s \nu_4^{[2]}
\right)\, ,  
\end{align}
where $\bp_A = (xh)^L\bP_A$ and
\begin{equation}
x\equiv x (u) = \frac{u+\sqrt{u^2 - 4 h^2}}{2 h}
\end{equation}
is the Zhukovsky variable used to parameterize the single cut of $\bP$ - functions on the defining Riemann sheet.

In summary, the equations we are going to solve are given by
\begin{align}
\frac{\nu_1^{[3]}}{\bP_1^{[1]}} - \frac{\nu_1^{[-1]}}{\bP_1^{[-1]}} - \s \left(
\frac{\bP_0^{[1]}}{\bP_1^{[1]}} - \frac{\bP_0^{[-1]}}{\bP_1^{[-1]}}
\right)\nu_1^{[1]} &= -\s \left(
\frac{\bP_2^{[1]}}{\bP_1^{[1]}} - \frac{\bP_2^{[-1]}}{\bP_1^{[-1]}}
\right)\nu_2^{[1]}\, , \label{Baxternu1} \\
\frac{\nu_2^{[3]}}{\bP_1^{[1]}} - \frac{\nu_2^{[-1]}}{\bP_1^{[-1]}} + \s \left(
\frac{\bP_0^{[1]}}{\bP_1^{[1]}} - \frac{\bP_0^{[-1]}}{\bP_1^{[-1]}}
\right)\nu_2^{[1]} &= \s \left(
\frac{\bP_3^{[1]}}{\bP_1^{[1]}} - \frac{\bP_3^{[-1]}}{\bP_1^{[-1]}}
\right)\nu_1^{[1]}\, , \label{Baxternu2}
\end{align}
and 
\begin{align}
\s \nu_1^{[2]} &= \bP_0 \nu_1 - \bP_2 \nu_2 + \bP_1 \nu_3 \, , \label{nu3sol} \\
\s \nu_2^{[2]} &= -\bP_0 \nu_2 + \bP_3 \nu_1 + \bP_1 \nu_4 \, , \label{nu4sol} \\
\btP_2 - \bP_2 &= \s \left(
\nu_3 \nu_1^{[2]} - \nu_1 \nu_3^{[2]}
\right)\, , \label{pt2} \\
\btP_1 - \bP_1 &= \s \left(
\nu_2 \nu_1^{[2]} - \nu_1 \nu_2^{[2]}
\right)\, , \label{pt1} \\
\left(
\nu_1 + \s \nu_1^{[2]}
\right) \left(
\bp_0 - (h x)^L
\right) &= \bp_2 \left(
\nu_2 + \s \nu_2^{[2]}
\right) - \bp_1 \left(
\nu_3 + \s \nu_3^{[2]}
\right)\, , \label{p0}
\\ 
\left(
\nu_2 + \s \nu_2^{[2]}
\right)\left(
\bp_0 + (h x)^L
\right) &= \bp_3 \left(
\nu_1 + \s \nu_1^{[2]}
\right) + \bp_1 \left(
\nu_4 + \s \nu_4^{[2]}
\right)\, .  \label{p3}
\end{align}
In addition, there are simple consequences of a given cut structure for $\nu$-functions, which will be used during solution. Namely, the following combinations of functions
\begin{align}
\nu_a (u) + \tilde{\nu}_a (u) &= \nu_a (u) + \s \nu_a^{[2]} (u)\, , \nonumber \\
\frac{\nu_a (u) - \tilde{\nu}_a (u)}{\sqrt{u^2 - 4 h^2}} &=
\frac{\nu_a (u) - \s \nu_a^{[2]} (u)}{\sqrt{u^2 - 4 h^2}}\, 
\label{cutsfree}
\end{align}
don't have cuts on the real axis.  To find a perturbative solution of the above system of equations we will use expansion of $\nu_a (u)$ functions in terms of QSC coupling constant $h$
\begin{equation}
\nu_{a} (u) = \sum_{l=0}^{\infty} h^{2 l - L} \nu_a^{(l)} (u)
\end{equation}
together with the following parametrization of $\bP$ - functions \cite{VolinPerturbativeSolution,ABJMQSC12loops} 
\begin{align}
\bP_1 &= (x h)^{-L} \bp_1 = (x h)^{-L} \left(
1+\sum_{k=1}^{\infty}\sum_{l=0}^{\infty} c_{1,k}^{(l)}\frac{h^{2 l+k}}{x^k}
\right)\, , \\
\bP_2 &= (x h)^{-L} \bp_2 = (x h)^{-L}\left(
\frac{h}{x} + \sum_{k=2}^{\infty}\sum_{l=0}^{\infty} c_{2,k}^{(l)} \frac{h^{2 l+k}}{x^k}
\right)\, , \\
\bP_0 &= (x h)^{-L} \bp_0  = (x h)^{-L} \left(
\sum_{l=0}^{\infty} A_0^{(l)} h^{2 l} u^L + 
\sum_{j=0}^{L-1}\sum_{l=0}^{\infty} m_j^{(l)} h^{2 l} u^j
+\sum_{k=1}^{\infty}\sum_{l=0}^{\infty} c_{0,k}^{(l)}
\frac{h^{2 l+k}}{x^k}
\right)\, , \\
\bP_3 &= (x h)^{-L} \bp_3 = (x h)^{-L}\left(
\sum_{l=0}^{\infty} A_3^{(l)} h^{2 l} u^{2 L+1}
+ \sum_{j=0}^{2 L}\sum_{l=0}^{\infty} k_j^{(l)} h^{2 l} u^j
+ \sum_{k=1}^{\infty}\sum_{l=0}^{\infty} c_{3,k}^{(l)}\frac{h^{2 l+k}}{x^k}
\right) \, .
\end{align}
where we have also accounted for large $u$ asymptotic of $\bP$ functions (\ref{Pasympt}).  We  would like to note, that, due to residual gauge symmetry of QSC equations,\footnote{See for details \cite{ABJMQSC12loops}.} the coefficients $m_j^{(l)}$, $k_j^{(l)}$ in the above parametrization of $\bP$ functions at twist $L=1$ are left undetermined. Next, the coefficients $A_0^{(l)}$, $A_3^{(l)}$ and $c_{i,k}^{(l)}$ are some functions of spin $S$ only and the mentioned gauge freedom can be also used to set $A_1 = 1$ and $A_2 = h^2$. The analytical continuation of $\bP$-functions through the cut on real axis is simple and is given by \cite{VolinPerturbativeSolution}:
\begin{equation}
\btP_a = \left(\frac{x}{h}\right)^L \btp_a\, , \quad \btp_a = \bp_a\Big|_{x\to 1/x}\, .
\end{equation}
In what follows we will consider perturbative solution in a special case of twist 1 operators, so from now on we put $L=1$.

\subsection{Sums of Baxter polynomials}\label{elementaryopSection}

Recently, in Ref. \cite{uspaceTMF} we have suggested that the full all-loop solution of the $\bP\mu$-system \eqref{nuPmonodromy} - \eqref{Pdiscont} can be obtained in terms of linear combinations of products of rational (in spectral parameter $u$), Hurwitz - functions and Baxter polynomials and showed an explicit example at four-loop order.  The purpose of this paper is to present explicit all-loop solution. To do that, let us first introduce the necessary notation.

The expressions for Baxter polynomials are obtained as leading order solutions for $\nu_1^{(0)} (u)$ - functions as follows. First, considering equations \eqref{Pasympt} and \eqref{P0} at large values of spectral parameter $u$ we get
\begin{equation}
A_0^{(0)}=i\sigma (2S+1)\,.
\end{equation} 
Substituting this expression into the first Baxter equation (\ref{Baxternu1}) and solving it in leading order in QSC coupling constant $h$ we have\footnote{One may use, for example, Mellin transform technique, see \cite{ABJM_QSC_Mellin} for more details.}
\begin{equation}
\nu_1^{(0)} (u) = \alpha Q_S^{[-1]} (u)\, ,
\end{equation} 
where $Q_S (u)$ is given by
\begin{equation}\label{eq:BaxterPolynomial}
Q_S\left(u\right) =\,_{2}F_{1}\left(-S,\tfrac{1}{2}+iu;1;2\right)
=\frac{(-)^{S}\Gamma\left(\frac{1}{2}+iu\right)}{S!\Gamma\left(\frac{1}{2}+iu-S\right)}\,_{2}F_{1}\left(-S,\tfrac{1}{2}+iu;\tfrac{1}{2}+iu-S;-1\right)\,.
\end{equation} 
and $\alpha$ is some spin-dependent constant to be determined later. 

Let us now introduce the following class of sums involving Baxter polynomials
\begin{align}
\left\langle Q\left(u\right)|w_{1}\left(\j\right),w_{2}\left(\j\right),\ldots,w_{n}\left(\j\right)\right\rangle  & =\sum_{S\geq j_{1}>j_{2}\ldots>j_{n}>0}Q_{S-j_{1}}\left(u\right)\prod_{k}w_{k}\left(j_{k}\right) , \label{QW-sums} \\
\left\langle Q\left(u\right)|\right\rangle  & =Q_{S}\left(u\right) , 
\end{align}
where $w_k$ are some weights. In the case, when the argument of $Q_S$ is $u$ we will often drop it and simply write $\left\langle Q|w_{1}\left(\j\right),w_{2}\left(\j\right),\ldots,w_{n}\left(\j\right)\right\rangle $.
We also introduce a shortcut
\begin{align}
\left\langle w_{1}\left(\j\right),w_{2}\left(\j\right),\ldots,w_{n}\left(\j\right)\right\rangle  & =\left\langle Q(\tfrac{i}{2})| w_{1}\left(\j\right),w_{2}\left(\j\right),\ldots,w_{n}\left(\j\right)\right\rangle\,.
=\sum_{S\geq j_{1}>j_{2}\ldots>j_{n}>0}\prod_{k}w_{k}\,.
\end{align} 
Note that the  $\left\langle w_{1},W\right\rangle$-sums satisfy usual stuffle relations.
Here and below we write weights $w_k(\j)$ in several equivalent ways, $w_k (\j) \equiv w_k (j) \equiv w_k$ and use $W$ to denote arbitrary (maybe empty) sequence of weights. 
In addition, we will use the notation
\begin{equation}\label{eq:ketsums}
\left|w_{1},w_{2},\ldots,w_{n}\right\rangle_{j_0} =\sum_{j_{0}>j_{1}\ldots>j_{n}>0}\prod_{k}w_{k}(j_k)\,,
\end{equation}
%\begin{equation}\label{eq:ketsums}
%\left|w_{1}\left(j_{1}\right),w_{2}\left(\j\right),\ldots,w_{n}\left(\j\right)\right\rangle =\sum_{j_{1}>j_{2}\ldots>j_{n}>0}\prod_{k}w_{k}\,,
%\end{equation}
so that
\[
\left\langle Q\left(u\right)|w_{1},w_{2},\ldots,w_{n}\right\rangle =\sum_{S\geq j_{1}>0}Q_{S-j_{1}}\left(u\right)w_{1}\left(j_{1}\right)\left|w_{2},\ldots,w_{n}\right\rangle_{j_1} \,.
\]
%\[
%\left\langle Q\left(u\right)|w_{1}\left(\j\right),w_{2}\left(\j\right),\ldots,w_{n}\left(\j\right)\right\rangle =\sum_{S\geq j_{1}>0}Q_{S-j_{1}}\left(u\right)\left|w_{1}\left(j_{1}\right),w_{2}\left(\j\right),\ldots,w_{n}\left(\j\right)\right\rangle \,.
%\]
It turns out that weights at twist 1 can always be reduced to a set of canonical weights for which we introduce special notations:
\begin{gather}\label{eq:cweights}
\frac{1}{\j^{n}}=n_{+}(\j)\,,\qquad\frac{\left(-\right)^{\j}}{\j^{n}}=n_{-}(\j)\,,\\
\frac{1}{\left(S+1-\j\right)^{n}}=\overline{n}_{+}(\j)\,,\qquad\frac{\left(-\right)^{\j}}{\left(S+1-\j\right)^{n}}=\overline{n}_{-}(\j)\,,\\
\frac{1}{\left(2S+1-\j\right)^{n}}=\hat{n}_{+}(\j)\,,\qquad\frac{\left(-\right)^{\j}}{\left(2S+1-\j\right)^{n}}=\hat{n}_{-}(\j)\,.
\end{gather}
In Ref. \cite{uspaceTMF} we have considered elementary operations on Baxter polynomials, such as shifts and partial fractions. The latter can be also extended to the sums of Baxter polynomials.  In particular, the shift in spectral parameter $u$ can be performed using
($a=\pm 1$):
\begin{equation}
Q_{S}^{\left[2a\right]}=Q_{S}+2\sum_{k=1}^{S}a^{k}Q_{S-k}=Q_{S}+2\left\langle Q|0_{a}\right\rangle 
\end{equation}
%TODO: choose symbol for sign: \eta? 
and 
\begin{equation}
\left\langle Q|w,W\right\rangle ^{\left[2a\right]}=\left\langle Q|w,W\right\rangle +2\left\langle Q|0_{a},0_{a}\cdot w,W\right\rangle\, . 
\end{equation} 
%TODO: explain W notation above 
%Next, for the multiplication by spectral parameter $u$ we have
%\begin{eqnarray}
%uQ_{S}=\frac{i}{2}\left[\left(S+1\right)Q_{S+1}-SQ_{S-1}\right]
%\end{eqnarray} 
Next, we have the rules for partial fractions ($a=\pm$):
\begin{align}
\frac{Q_{S}}{u+a\frac{i}{2}} & =\frac{\left(-a\right)^{S}}{u+a\frac{i}{2}}+2ia\left\langle Q|0_{a},\overline{1}_{-}\right\rangle +2ia\left\langle Q|\overline{1}_{-a}\right\rangle 
\end{align}
and
\begin{align}
\frac{\left\langle Q|w,W\right\rangle }{u+a\frac{i}{2}} & =\frac{\left(-a\right)^{S}}{u+a\frac{i}{2}}\left\langle 0_{-a}\cdot w,W\right\rangle +2ia\left\langle Q|0_{a},\overline{1}_{-},0_{-a}\cdot w,W\right\rangle +2ia\left\langle Q|\overline{1}_{-a},0_{-a}\cdot w,W\right\rangle\,. 
\end{align}
Finally we can shift the spin $S$ of the Baxter polynomials using
\begin{align}\label{eq:Sshift}
\left(S+1\right)\begin{pmatrix}Q_{S+1}^{\left[1\right]}\\
Q_{S+1}^{\left[-1\right]}
\end{pmatrix} & =\begin{pmatrix}S+1-iu & -iu\\
-iu & -S-1-iu
\end{pmatrix}\begin{pmatrix}Q_{S}^{\left[1\right]}\\
Q_{S}^{\left[-1\right]}
\end{pmatrix}\,,\\
S\begin{pmatrix}Q_{S-1}^{\left[1\right]}\\
Q_{S-1}^{\left[-1\right]}
\end{pmatrix} & =\begin{pmatrix}S+iu & -iu\\
-iu & -S+iu
\end{pmatrix}\begin{pmatrix}Q_{S}^{\left[1\right]}\\
Q_{S}^{\left[-1\right]}
\end{pmatrix}\,.
\end{align} 
Remarkably, the introduced class of functions, $\langle Q|\ldots\rangle$, is closed under differentiation. In order to prove this, let us first consider the sums
\begin{align}\label{eq:summod}
\left|\tfrac1{j_0-j_1}w_{1}\left(j_{1}\right),W\right\rangle_{j_0}
=\sum_{j_0>j_{1}>j_{2}\ldots>j_{n}>0}\frac{w_1(j_1)}{j_0-j_1}w_{2}(j_2)\ldots w_{n}(j_n)\,.
\end{align}
Let us prove that these sums reduce to the linear combination of our standard sums \eqref{eq:ketsums}. We proceed by induction over the depth of the sum. Let us first write
\begin{equation}
\frac{w_1(j_1)}{j_0-j_1}=
\frac{w_1(j_1)-(\pm)^{j_0-j_1}w_1(j_0)}{j_0-j_1}+
(\pm)^{j_0}w_1(j_0)\frac{(\pm)^{j_1}}{j_0-j_1}\,,
\end{equation}
where the lower sign is chosen if $w_1(j_1)$ contains the factor $(-)^{j_1}$. Then the denominator $j_0-j_1$ cancels in the first term and, therefore, this term gives rise only to standard sums \eqref{eq:ketsums}. The second term gives rise to the sums
\begin{align}\label{eq:summod1}
\left|\tfrac{(\pm)^{j}}{j_0-j},W\right\rangle_{j_0}=\sum_{j_0>j_{1}\ldots>j_{n}>0}\frac{(\pm)^{j_1}}{j_0-j_1}w_{2}(j_2)\ldots w_{n}(j_n)
\end{align}
In order to transform these sums, we observe that
\begin{equation}
\sum_{j_{1}=j_2+1}^{j_0-1}\frac{(\pm)^{j_1}}{j_0-j_1}=\sum_{j_{1}=j_2+1}^{j_0-1}\frac{(\pm)^{j_0+j_2-j_1}}{j_1-j_2}\,.
\end{equation}
This identity is  proved by the substitution $j_1\to j_0+j_2-j_1$. Then
\begin{align}\label{eq:summod2}
\sum_{j_0>j_{1}>j_{2}\ldots>j_{n}>0}\frac{(\pm)^{j_1}}{j_0-j_1}w_{2}(j_2)\ldots w_{n}(j_n)
=\sum_{j_{1}=1}^{j_0}(\pm)^{j_0+j_1}\sum_{j_{1}>j_{2}\ldots>j_{n}>0}\frac{(\pm)^{j_2}}{j_1-j_2}w_{2}(j_2)\ldots w_{n}(j_n)\,,
\end{align}
and the inner sum is again of the form \eqref{eq:summod}, but the depth is reduced. This proves the induction step.

Now, using the differentiation formula\footnote{It can be obtained using generating function for Baxter polynomials and we leave the proof of this formula to the reader.}
\begin{equation}
	i\partial_u Q_S =\sum_{j_0=1}^{S} \frac{Q_{S-j_0}}{j_0}\left(1-(-)^{j_0}\right)=\langle Q|1_+\rangle-\langle Q|1_-\rangle\,.
\end{equation}
we have
\begin{multline}
i\partial_u \langle Q|w,W\rangle =\sum_{j_1=1}^{S}i\partial_u Q_{S-j_1} |w(j_1),W\rangle=\sum_{j_1=1}^{S}\sum_{j_0=1}^{S-j_1} \frac{Q_{S-j_1-j_0}}{j_0}\left(1-(-)^{j_0}\right) |w(j_1),W\rangle\\
=\sum_{j_1=1}^{S}\sum_{j_0=j_1+1}^{S} \frac{Q_{S-j_0}}{j_0-j_1}\left(1-(-)^{j_0+j_1}\right) |w(j_1),W\rangle
=\sum_{j_0=1}^{S} Q_{S-j_0}\sum_{j_1=1}^{j_0-1}\frac{1-(-)^{j_0+j_1}}{j_0-j_1}|w(j_1),W\rangle\\
=\sum_{j_0=1}^{S} Q_{S-j_0}\sum_{j_1=1}^{j_0-1}\frac{1}{j_0-j_1}|w(j_1),W\rangle
-\sum_{j_0=1}^{S} Q_{S-j_0}(-)^{j_0}\sum_{j_1=1}^{j_0-1}\frac{(-)^{j_1}}{j_0-j_1}|w(j_1),W\rangle
\end{multline}
Since the inner sums in the last expression are both of the form \eqref{eq:summod}, they can be expressed as a linear combination of the standard sums \eqref{eq:ketsums}. Therefore, $i\partial_u \langle Q|\ldots \rangle$ can indeed be expressed as a linear combination of $\langle Q|\ldots \rangle$ sums. 
In particular, the expansion of $\langle Q|W\rangle$-sums  at $u=i/2$ can be expressed  
in terms of  $\left\langle W\right\rangle$-sums.

Let us summarize the results of the present subsection.  We introduced the class of functions - products of rational functions in spectral parameter $u$ with $\langle Q|W\rangle$-sums \eqref{QW-sums} closed under elementary operations, such as argument shifts and partial fractions, as well as under differentiation. As we will see in next subsection this class of functions extended to products with Hurwitz functions is also sufficient for finding a perturbative solution of inhomogeneous Baxter equations.  
 
\subsection{Solutions of Baxter equations}

The most complicated part in the perturbative solution of Riemann-Hilbert problem for $\bP\nu$-system is the solution of two inhomogeneous Baxter equations  \eqref{Baxternu1} and \eqref{Baxternu2}. Iterative perturbative solution of equations \eqref{Baxternu1}--\eqref{p3} goes through the expansion in QSC coupling constant $h$.  In particular we expand equations \eqref{Baxternu1}, \eqref{Baxternu2} up to $h^k$ and obtain the following inhomogeneous Baxter equations for $q_{1,2}^{(k)}=\left(\nu_{1,2}^{(k)}\right)^{[1]}$
\begin{align}
\B_1\left(q_{1}^{(k)}\right)\equiv (u+i/2)q_{1}^{(k)}(u+i)-i(2S+1)q_{1}^{(k)}(u)-(u-i/2)q_{1}^{(k)}(u-i)=&V_{1}^{(k)} \label{Baxter1a}\,,\\
\B_{-1}\left(q_{2}^{(k)}\right) \equiv(u+i/2)q_{2}^{(k)}(u+i)+ i(2S+1)q_{2}^{(k)}(u)-(u-i/2)q_{2}^{(k)}(u-i)=&V_{2}^{(k)}\,, \label{Baxter2a}
\end{align}
where $V_{1}^{(k)}$ depends on $q_{1,2}^{(l)}$ with $l<k$, and $V_{2}^{(k)}$ depends in addition on $q_1^{(k)}$. 
The solution of these equations contains in general two pieces: the solution of homogeneous equation with arbitrary periodic coefficients and some particular solution of nonhomogeneous equation.  

\subsubsection{Homogeneous solution}

The first homogeneous solutions of Baxter equations \eqref{Baxter1a} and \eqref{Baxter2a} are easy to find, they are given  by\footnote{See the discussion at the beginning of subsection \ref{elementaryopSection}.} $\Phi_Q^{\mathrm{per}} (u) Q_S (u)$ and $\Phi_Q^{\mathrm{anti}} (u) Q_S (u)$ correspondingly. Here $\Phi_Q^{\mathrm{per}} (u)$ and $\Phi_Q^{\mathrm{anti}} (u)$ are arbitrary $i$-(anti)periodic functions of spectral parameter $u$. To find second solutions let us consider the following identity
\begin{equation}
\mathcal{B}_{-1}\left(\xi_{-1}Q_{S}\right)=Q_{S}^{\left[2\right]}-Q_{S}^{\left[-2\right]}=2\sum_{k=1}^{S}\left[1-\left(-1\right)^{k}\right]Q_{S-k}=-i\sum_{k=1}^{S}\frac{1-\left(-1\right)^{k}}{2S-k+1}\mathcal{B}_{-1}Q_{S-k}\,,
\end{equation}
where\footnote{See appendix \ref{Hurwitz-functions} for full definition of Hurwitz functions $\xi_A$.} $\xi_{-1}(u)=-\sum_{n=1}^{\infty}\tfrac{(-)^n}{u+in-\frac i2}$.
It is easy to see that the second homogeneous solution of the second Baxter equation \eqref{Baxter2a} is given by
\begin{equation}
\Z_S (u) = \xi_{-1}Q_{S}+i\sum_{k=1}^{S}\frac{1-\left(-1\right)^{k}}{2S-k+1}Q_{S-k}
=\xi_{-1}Q_{S}+i\langle Q|\hat{1}_+\rangle-i\langle Q|\hat{1}_-\rangle\,.
\end{equation}
Then the general solutions of first and second homogeneous Baxter equations are given by
\begin{align}
q_1^{\mathrm{hom}} (S,u) &= \Phi_{1}^{\mathrm{per}} Q_S(u) + \Phi_{1}^{\mathrm{anti}} \Z_S (u)\, , \\
q_2^{\mathrm{hom}} (S,u) &= \Phi_{2}^{\mathrm{anti}} Q_S(u) + \Phi_{2}^{\mathrm{per}} \Z_S (u)\, ,
\end{align}
where $\Phi_{a}^{\mathrm{per}}$ and $\Phi_{a}^{\mathrm{anti}}$  are arbitrary $i$-periodic and $i$-anti-periodic functions in spectral parameter $u$. They have to be determined from the consistency conditions implied by the equations \eqref{Baxternu1}-\eqref{cutsfree}. We will parametrize their $u$ dependence similar to Ref. \cite{VolinPerturbativeSolution,ABJMQSC12loops} using the basis of $i$-periodic and $i$-anti-periodic combinations of Hurwitz functions defined as 
%\begin{equation}
%\P_k (u) = \eta_k (u) + \text{sgn}(k) \bar{\eta}_k^{[-2]} (u) = \P_k (u+i)\, , \quad k\neq 0 \in \mathbb{Z}\, ,
%\end{equation}
\begin{equation}
\P_k (u) = \xi_k^{[-1]}(u) + \text{sgn}(k)(-)^k \xi_k^{[-3]}(-u) = \text{sgn}(k)\P_k (u+i)\, , \quad k\neq 0 \in \mathbb{Z}\, ,
\end{equation}
Note that $\P_k (u)$ can be expressed via elementary functions:
\begin{equation}
\P_k (u) = \frac{(-\partial_u)^{|k|-1}}{(|k|-1)!} \begin{cases} 
\pi\coth(\pi u) & k > 0 \\
\pi/\cosh(\pi u)  & k < 0
\end{cases}
\end{equation}
Then the functions  $\Phi_{a}^{\mathrm{per}}$ and $\Phi_{a}^{\mathrm{anti}}$ are written as
\begin{equation}
\Phi_{a}^{\mathrm{per}} (u) = \phi_{a,0}^{\mathrm{per}} + \sum_{j=1}^{\Lambda_a^{\mathrm{per}}}\phi_{a,j}^{\mathrm{per}} \P_j (u+i/2)\, , \quad \Phi_{a}^{\mathrm{anti}} (u) =  \sum_{j=1}^{\Lambda_a^{\mathrm{anti}}}\phi_{a,j-1}^{\mathrm{anti}}\P_{-j} (u+i/2)
\end{equation}
where the upper limits of summation depend on the order of perturbation theory as follows
\begin{align}
\Lambda_1^{(k),\mathrm{per}} &= \Lambda_2^{(k),\mathrm{anti}} = 2k-1\, , \\
\Lambda_1^{(k),\mathrm{anti}} &= \Lambda_2^{(k),\mathrm{per}} = 2k-2\, .
\end{align}
Here $k = 1$ for NLO, $k=2$ for NNLO, and so on.

\subsubsection{Dictionary for inhomogeneous solutions}

To find particular solutions of Baxter equations \eqref{Baxter1a} and \eqref{Baxter2a} let us introduce the operators $\mathcal{F}_{\pm 1}$ which are right inverse of the Baxter operators $\mathcal{B}_{\pm 1}$, Eqs. \eqref{Baxter1a}, \eqref{Baxter2a}, i.e., satisfy
\begin{equation}
\mathcal{B}_{\pm 1}(\mathcal{F}_{\pm 1}(f))=f\,.
\end{equation}
Note that they are nothing but the operators $\mathcal{F}_{1,2}^{S}$ introduced in our previous paper \cite{uspaceTMF}. Our basic idea now is to compile a dictionary sufficient to treat all the functions appearing in the right-hand sides of Baxter equations. 

\paragraph{Action of $\FF_{\pm1}$ on $\left\langle Q|W\right\rangle$.}~\\
We first act with the operators $\mathcal{B}_{\pm 1}$ on the functions $\langle Q\left(u\right)|w,W\rangle$:

\begin{align}
\label{eq:B+_Q}
	\B_{1}\left[\left\langle Q|w,W\right\rangle \right]	&=\sum_{j=1}^{S}\B_{1}\left[Q_{S-j}\right]w\left(j\right)\left|W\right\rangle_j \\ &=-2i\sum_{j=1}^{S}Q_{S-j}jw\left(j\right)\left|W\right\rangle\nonumber_j
	=\left\langle Q|(-1)_+\cdot w,W\right\rangle\,,\\
\label{eq:B-_Q}
	\B_{-1}\left[\left\langle Q|w,W\right\rangle \right]&=\sum_{j=1}^{S}\B_{-1}\left[Q_{S-j}\right]w\left(j\right)\left|W\right\rangle_j \\
	& =2i\sum_{j=1}^{S}Q_{S-j}\left(2S-j+1\right)w\left(j\right)\left|W\right\rangle\nonumber_j=\left\langle Q|(\widehat{-1})_+\cdot w,W\right\rangle
\end{align}
Replacing $w\to 1_+\cdot w$ in the first formula and $w\to \hat{1}_+ \cdot w$ in the second we obtain the following entries in our dictionary:
\begin{align}
\label{eq:FVQ1}
\FF_{1}\left[\left\langle Q|w,W\right\rangle \right]=\tfrac{i}{2}\left\langle Q|1_+\cdot w,W\right\rangle\,,\\
\label{eq:FVQ2}\FF_{-1}\left[\left\langle Q|w,W\right\rangle \right]=-\tfrac{i}{2}\left\langle Q|\hat{1}_+\cdot w,W\right\rangle\,.
\end{align}
Special cases $\FF_{\pm1}\left[\langle Q|\rangle \right]$ can be read off from Eqs. (3.42) and (3.43) of Ref. \cite{uspaceTMF}:
\begin{align}
\FF_1 \left[\langle Q|\rangle\right] &= -\tfrac{1}{2}\langle Q|\rangle\,  \xi_1 - \tfrac{i}{2}\langle Q|1_+\rangle- \tfrac{i}{2}\langle Q|1_-\rangle\,,\\
\FF_{-1} \left[\langle Q|\rangle\right] &=  -\tfrac{i}{2(2S + 1)}\langle Q|\rangle\, .
\end{align}

\paragraph{Action of $\FF_{\pm1}$ on $\left\langle Q|W\right\rangle\xi_{a_1,a_2,\ldots}$.}~\\
The basic idea of calculating $\FF_{\pm1}[\left\langle Q|W\right\rangle\xi_{A}]$ is to use the analogue of summation-by-part formulae from Ref. \cite{VolinPerturbativeSolution,ABJMQSC12loops}. For any two functions $f$ and $g$ we have\footnote{In the present section $\sigma = \pm1$ is an arbitrary sign, not to be confused with $\sigma=(-)^S$ entering \eqref{nuanalytcont} and other equations for $\bP\nu$-system.}
\begin{multline}
	\B_{\sigma}\left[fg\right]=f\B_{\sigma}\left[g\right]+\left(u+\tfrac{i}{2}\right)\left(f^{\left[2\right]}-f\right)g^{\left[2\right]}-\left(u-\tfrac{i}{2}\right)\left(f^{\left[-2\right]}-f\right)g^{\left[-2\right]}\\
	=-f\B_{-\sigma}\left[g\right]+\left(u+\tfrac{i}{2}\right)\left(f^{\left[2\right]}+f\right)g^{\left[2\right]}-\left(u-\tfrac{i}{2}\right)\left(f^{\left[-2\right]}+f\right)g^{\left[-2\right]}\,.
\end{multline}
Substituting $g\to \FF_{\pm\sigma}[g]$ and $f= \xi_{a,A}$, we obtain 
\begin{multline}\label{eq:Fgxi}
	\FF_{\sigma}\left[g \xi_{\sigma_a|a|,A} \right]=\sigma_a\xi_{\sigma_a|a|,A} \FF_{\sigma_a\sigma}\left[g\right]
	-\sigma_a \FF_{\sigma}\bigg[\left(u+\tfrac{i}{2}\right)\left(\xi_{\sigma_a|a|,A} ^{\left[2\right]}-\sigma_a\xi_{\sigma_a|a|,A} \right)\FF_{\sigma_a\sigma}\left[g\right]^{\left[2\right]}\\
	-\left(u-\tfrac{i}{2}\right)\left(\xi_{\sigma_a|a|,A}^{\left[-2\right]} -\sigma_a\xi_{\sigma_a|a|,A} \right)\FF_{\sigma_a\sigma}\left[g\right]^{\left[-2\right]}\bigg]\,.
\end{multline}
Here $A=a_1,a_2\ldots$ stands for an arbitrary (maybe empty) sequence of indices, and $\sigma_a=a/|a|$.
Then, for $g=\langle Q|W\rangle$, using Eqs. \eqref{eq:FVQ1}, \eqref{eq:FVQ2}, and \eqref{eq:xishift}, we have

	%Similarly, we act with $\B_{\pm 1}$ on $\langle Q|w,W\rangle \xi_{a,A}$.
%We have
%\begin{multline}\label{eq:BVQxi}
%\B_{\sigma}\left[\langle Q|w,W\rangle\xi_{a,A}\right] =\left(u+\tfrac{i}{2}\right)\langle Q|w,W\rangle^{\left[2\right]}\left(\sigma_{a}\xi_{a,A}-\frac{\sigma_{a}}{\left(u+\frac{i}{2}\right)^{\left|a\right|}}\xi_{A}^{\left[2\right]}\right)\\
%-\sigma\left(2S+1\right)\langle Q|w,W\rangle\xi_{a,A}
%-\left(u-\tfrac{i}{2}\right)\langle Q|w,W\rangle^{\left[-2\right]}\left(\sigma_{a}\xi_{a,A}+\frac{1}{\left(u-\frac{i}{2}\right)^{\left|a\right|}}\xi_{A}\right)\\
% =\sigma_{a}\B_{\sigma\sigma_{a}}\left[\langle Q|w,W\rangle\right]\xi_{a,A}-\frac{\sigma_{a}}{\left(u+\frac{i}{2}\right)^{\left|a\right|-1}}\langle Q|w,W\rangle^{\left[2\right]}\xi_{A}^{\left[2\right]}-\frac{1}{\left(u-\frac{i}{2}\right)^{\left|a\right|-1}}\langle Q|w,W\rangle^{\left[-2\right]}\xi_{A}\,,
%\end{multline}
%where $\sigma = \pm 1$ and $\sigma_a=\mathrm{sgn}\,a$. Now, if $\sigma\sigma_a = 1$ ($\sigma\sigma_a = -1$), we use Eq. \eqref{eq:B+_Q} (Eq. \eqref{eq:B-_Q}) and make the substitution $w\to 1_+\cdot w$ ($w\to \hat{1}_+\cdot w$) so that $\B_{\sigma\sigma_{a}}\left[\langle Q|w,W\rangle\right]\to \langle Q|w,W\rangle$ in the right-hand side. 

\begin{align}
\FF_{\sigma}\left[\left\langle Q|w,W\right\rangle \xi_{\sigma\left|a\right|,A}\right] & =\frac{i}{2}\bigg\{\sigma \left\langle Q|1_{+}\cdot w,W\right\rangle \xi_{\sigma\left|a\right|,A}\nonumber\\
&+\FF_{\sigma}\bigg[\frac{\left\langle Q|1_{+}\cdot w,W\right\rangle ^{\left[2\right]}\xi_{A}^{\left[2\right]}}{\left(u+\frac{i}{2}\right)^{\left|a\right|-1}}+\frac{\sigma\left\langle Q|1_{+}\cdot w,W\right\rangle ^{\left[-2\right]}\xi_{A}}{\left(u-\frac{i}{2}\right)^{\left|a\right|-1}}\bigg]\bigg\} \\
\FF_{\sigma}\left[\left\langle Q|w,W\right\rangle \xi_{-\sigma\left|a\right|,A}\right] & =\frac{i}{2}\bigg\{\sigma \left\langle Q|\hat{1}_{+}\cdot w,W\right\rangle \xi_{-\sigma\left|a\right|,A}\nonumber\\
&+\FF_{\sigma}\bigg[-\frac{\left\langle Q|\hat{1}_{+}\cdot w,W\right\rangle ^{\left[2\right]}\xi_{A}^{\left[2\right]}}{\left(u+\frac{i}{2}\right)^{\left|a\right|-1}}+\frac{\sigma\left\langle Q|\hat{1}_{+}\cdot w,W\right\rangle ^{\left[-2\right]}\xi_{A}}{\left(u-\frac{i}{2}\right)^{\left|a\right|-1}}\bigg]\bigg\}\,.
%\\\FF_{-1}\left[\left\langle Q|w,W\right\rangle \xi_{-\left|a\right|,A}\right] & =-\frac{i}{2}\bigg\{ \left\langle Q|1_{+}\cdot w,W\right\rangle \xi_{-\left|a\right|,A}\nonumber\\
%&+\FF_{-1}\bigg[-\frac{\left\langle Q|1_{+}\cdot w,W\right\rangle ^{\left[2\right]}\xi_{A}^{\left[2\right]}}{\left(u+\frac{i}{2}\right)^{\left|a\right|-1}}+\frac{\left\langle Q|1_{+}\cdot w,W\right\rangle ^{\left[-2\right]}\xi_{A}}{\left(u-\frac{i}{2}\right)^{\left|a\right|-1}}\bigg]\bigg\}\,, \\
%\FF_{-1}\left[\left\langle Q|w,W\right\rangle \xi_{\left|a\right|,A}\right] & =-\frac{i}{2}\bigg\{ \left\langle Q|\hat{1}_{+}\cdot w,W\right\rangle \xi_{\left|a\right|,A}\nonumber\\
%&+\FF_{-1}\bigg[\frac{\left\langle Q|\hat{1}_{+}\cdot w,W\right\rangle ^{\left[2\right]}\xi_{A}^{\left[2\right]}}{\left(u+\frac{i}{2}\right)^{\left|a\right|-1}}+\frac{\left\langle Q|\hat{1}_{+}\cdot w,W\right\rangle ^{\left[-2\right]}\xi_{A}}{\left(u-\frac{i}{2}\right)^{\left|a\right|-1}}\bigg]\bigg\}\,. 
\end{align}
The operators $\FF_{\pm1}$ in the right-hand side of the above equations act on the `simpler' objects because the number of indices of $\xi$-functions is reduced by one.

Let us present separately the formula
\begin{align}
\FF_{\sigma}\left[\langle Q|\rangle\xi_{-\sigma\left|a\right|,A}\right] & =\frac{i}{2\left(2S+1\right)}\left\{\sigma \langle Q|\rangle\xi_{-\left|a\right|,A}+\FF_{1}\left[-\frac{\langle Q|\rangle^{[2]}\xi_{A}^{\left[2\right]}}{\left(u+\frac{i}{2}\right)^{\left|a\right|-1}}+\frac{\sigma\langle Q|\rangle^{\left[-2\right]}\xi_{A}}{\left(u-\frac{i}{2}\right)^{\left|a\right|-1}}\right]\right\}
%\,,\\
%\FF_{-1}\left[\langle Q|\rangle\xi_{\left|a\right|,A}\right] & =-\frac{i}{2\left(2S+1\right)}\left\{ \langle Q|\rangle\xi_{\left|a\right|,A}+\FF_{-1}\left[\frac{\langle Q|\rangle^{\left[2\right]}\xi_{A}^{\left[2\right]}}{\left(u+\frac{i}{2}\right)^{\left|a\right|-1}}+\frac{\langle Q|\rangle^{\left[-2\right]}\xi_{A}}{\left(u-\frac{i}{2}\right)^{\left|a\right|-1}}\right]\right\}
\,. 
\end{align}
In order to find $\FF_{\sigma}\left[\langle Q|\rangle\xi_{\sigma\left|a\right|,A}\right]$, we consider the identity
\begin{equation}
	\B_{\sigma}\left[\langle Q|\rangle\xi_{\sigma,\sigma|a|,A}\right]=-\sigma\langle Q|\rangle^{\left[2\right]}\xi_{\sigma|a|,A}^{\left[2\right]}-\langle Q|\rangle^{\left[-2\right]}\xi_{\sigma|a|,A}\,.
\end{equation}
We transform the right-hand side using the formulas \eqref{eq:xishift} and obtain
\begin{equation}
	\B_{\sigma}\left[\langle Q|\rangle\xi_{\sigma,\sigma|a|,A}\right]=-2\langle Q|\rangle\xi_{\sigma|a|,A}+\frac{\langle Q|\rangle^{\left[2\right]}\xi_{A}^{\left[2\right]}}{\left(u+\frac{i}{2}\right)^{|a|}}
	-2\left\langle Q|0_{+}\right\rangle \xi_{\sigma|a|,A}-2\left\langle Q|0_{-}\right\rangle \xi_{\sigma|a|,A}\,.
\end{equation}
Then we obtain
\begin{equation}
\FF_{\sigma}\left[\langle Q|\rangle\xi_{\sigma|a|,A}\right]=-\frac12\Bigg\{\langle Q|\rangle\xi_{\sigma,\sigma|a|,A}
+\FF_{\sigma}\left[\frac{\langle Q|\rangle^{\left[2\right]}\xi_{A}^{\left[2\right]}}{\left(u+\frac{i}{2}\right)^{|a|}}
-2\left\langle Q|0_{+}\right\rangle \xi_{\sigma|a|,A}-2\left\langle Q|0_{-}\right\rangle \xi_{\sigma|a|,A}\right]\Bigg\}\,.
\end{equation}

\paragraph{Action of $F_{\pm1}$ on $\frac{\xi_A}{\left(u\pm i/2\right)^{a}}$}~\\
From now on we will present only the final entries of our dictionary as the derivations goes more or less along the same lines as before.
The action of $\FF_{\pm1}$ on $\frac{\xi_A}{\left(u\pm i/2\right)^{a}}$ is the following
\begin{multline}
	\FF_{\sigma}\left[\frac{1}{\left(u+\frac{i}{2}\right)^{\left|a\right|}}\xi_{b,A}\right]=
	-\sigma_{b}\langle Q|\rangle\xi_{\sigma,-\sigma\left|a\right|,b,A}\\
	+\FF_{\sigma}\left[\frac{\langle Q|\rangle^{\left[2\right]}}{\left(u+\frac{i}{2}\right)^{\left|a\right|+\left|b\right|}}\xi_{A}^{\left[2\right]}-\frac{\langle Q|\rangle^{\left[2\right]}-1}{\left(u+\frac{i}{2}\right)^{\left|a\right|}}\xi_{b,A}+2\left[\left\langle Q|0_{+}\right\rangle -\left\langle Q|0_{-}\right\rangle \right]\sigma_{b}\xi_{-\sigma\left|a\right|,b,A}\right]\,.
\end{multline}

\begin{multline}
\FF_{\sigma}\left[\frac{1}{\left(u+\frac{i}{2}\right)^{\left|a\right|}}\right]=-\langle Q|\rangle\xi_{\sigma_,-\sigma\left|a\right|}
+\FF_{\sigma}\left[-\frac{\langle Q|\rangle^{\left[2\right]}-1}{\left(u+\frac{i}{2}\right)^{\left|a\right|}}+2\left[\left\langle Q|0_{+}\right\rangle -\left\langle Q|0_{-}\right\rangle \right]\xi_{-\sigma\left|a\right|}\right]\,.
\end{multline}

\begin{multline}
\FF_{\sigma}\left[\frac{\xi_{A}}{\left(u-\frac{i}{2}\right)^{\left|a\right|}}\right]
=\left(-\right)^{S}\Bigg\{\langle Q|\rangle\left(\sigma\xi_{\sigma,-\sigma\left|a\right|,A}-\xi_{\sigma\left|a\right|+\sigma,A}\right)\\
+\FF_{\sigma}\left[-\sigma\frac{\langle Q|\rangle^{\left[-2\right]}-\left(-\right)^{S}}{\left(u-\frac{i}{2}\right)^{\left|a\right|}}\xi_{A}+2\left[\left\langle Q|0_{-}\right\rangle -\left\langle Q|0_{+}\right\rangle \right]\xi_{-\sigma\left|a\right|,A}\right]\Bigg\}\,.
\end{multline}

\begin{multline}
\FF_{\sigma}\left[\frac{1}{\left(u-\frac{i}{2}\right)^{\left|a\right|}}\right]
=\left(-\right)^{S}\Bigg\{\langle Q|\rangle\left(\sigma\xi_{\sigma,-\sigma\left|a\right|}-\xi_{\sigma\left|a\right|+\sigma}\right)\\
+\FF_{\sigma}\left[-\sigma\frac{\langle Q|\rangle^{\left[-2\right]}-\left(-\right)^{S}}{\left(u-\frac{i}{2}\right)^{\left|a\right|}}+2\left[\left\langle Q|0_{-}\right\rangle -\left\langle Q|0_{+}\right\rangle \right]\xi_{-\sigma\left|a\right|}\right]\Bigg\}\,.
\end{multline}

\paragraph{Action of $F_{\pm1}$ on $u^n\langle Q|W\rangle {\xi_A}$}~\\
Finally, the right-hand side of the Baxter equations may also contain terms of the form $u^n\langle Q|W\rangle {\xi_A}$ with $n=1,2$. First, we use the same summation-by-part technique as before. Namely, we use Eq. \eqref{eq:Fgxi} with $g=u^n\langle Q|W\rangle$ to reduce the problem to the calculation of $\FF_\sigma[u^n\langle Q|W\rangle]$.

In order to calculate $\FF_\sigma[u^n\langle Q|W\rangle]$, it is convenient to introduce notation
\begin{equation}\label{eq:calQdef}
	\mathcal{Q}_{S}^{\left\{ n\right\} }	=\left[\frac{\Gamma\left(S+1+n\right)}{\Gamma\left(S+1\right)}\right]^{\mathrm{sgn}\left(n\right)}Q_{S+n}=\begin{cases}
	\left(S+1\right)_{n}Q_{S+n}, & n\geq0\\
	\left(S+1+n\right)_{-n}Q_{S+n}, & n<0
	\end{cases}
\end{equation} 
In terms of these functions we can easily express the required action:
\begin{align}\label{eq:calQinrhs}
	\FF_{-1}\left[u\left\langle Q|w(j),W\right\rangle \right]&=\frac{1}{4}\left\langle \mathcal{Q}^{\left\{ 1\right\} }|\tfrac{w(j)}{2S-j+2},W\right\rangle -\frac{1}{4}\left\langle \mathcal{Q}^{\left\{ -1\right\} }|\tfrac{w(j)}{2S-j},W\right\rangle\,, \\
	\FF_{1}\left[u\left\langle Q|w(j),W\right\rangle \right]&=-\frac{1}{4}\left\langle \mathcal{Q}^{\left\{ 1\right\} }|\tfrac{w(j)}{j-1},W\right\rangle +\frac{1}{4}\left\langle \mathcal{Q}^{\left\{ -1\right\} }|\tfrac{w(j)}{j+1},W\right\rangle \,,\\
	\FF_{-1}\left[u^{2}\left\langle Q|w(j),W\right\rangle \right]&=\frac{i}{8}\left\langle \mathcal{Q}^{\left\{ 2\right\} }|\tfrac{w(j)}{2S-j+3},W\right\rangle +\frac{i}{8}\left\langle \mathcal{Q}^{\left\{ -2\right\} }|\tfrac{w(j)}{2S-j-1},W\right\rangle 
	\nonumber\\
	& -\frac{i}{8}\left\langle Q|\tfrac{\left(S-j+1\right)^{2}+\left(S-j\right)^{2}}{2S-j+1}w(j),W\right\rangle\,,\\
	\FF_{1}\left[u^{2}\left\langle Q|w(j),W\right\rangle \right]&=-\frac{i}{8}\left\langle \mathcal{Q}^{\left\{ 2\right\} }|\tfrac{w(j)}{j-2},W\right\rangle -\frac{i}{8}\left\langle \mathcal{Q}^{\left\{ -2\right\} }|\tfrac{w(j)}{j+2},W\right\rangle
	\nonumber\\
	& +\frac{i}{8}\left\langle Q|\tfrac{\left(S-j+1\right)^{2}+\left(S-j\right)^{2}}{j}w(j),W\right\rangle\,.
\end{align}
Here
\begin{equation}
\left\langle \mathcal{Q}^{\left\{n\right\} }|w_{1}\left(\j\right),w_{2}\left(\j\right),\ldots,w_{n}\left(\j\right)\right\rangle = \sum_{S\geq j_{1}>j_{2}\ldots>j_{n}>0}\mathcal{Q}^{\left\{n\right\} }_{S-j_{1}}\left(u\right)\prod_{k}w_{k}\left(j_{k}\right)
\end{equation}
and we assume that $w(j)$ is one of the canonical weights \eqref{eq:cweights}.
Then we use the partial-fractioning identities similar to 
\begin{equation}\label{eq:pf}
	\frac{w(j)}{j-1}=\frac{w(j) - (\pm)^{j-1}w(1)}{j-1}+w(1)\frac{(\pm)^{j-1}}{j-1}\,,
\end{equation}
where we choose the lower sign if $w(j)$ contains $(-)^j$ factor and the upper sign otherwise. Then the first term in Eq. \eqref{eq:pf} is obviously a combination of canonical weights. The second term contains a shifted weight $1_\pm (j-1)$. Note that this shift is correlated with  superscript $\{n\}$ of $\mathcal{Q}$ functions. Therefore, we need the transformation rules for the sums of the form
\begin{equation}\label{eq:calQ1}
	\left\langle \mathcal{Q}^{\left\{n\right\} }|w\left(\j\right),W\right\rangle,
\end{equation} 
where  $w\left(\j\right)$ is one of the canonical weights \eqref{eq:cweights},
 and 
\begin{equation}\label{eq:calQ2}
\left\langle \mathcal{Q}^{\left\{n\right\} }|\tilde{1}\left(\j-n\right),W\right\rangle\,,
\end{equation} 
where $\tilde{1}$ is one of the four weights $1_+,\ 1_-,\ \hat{1}_+,\ \hat{1}_-$.
First, we note simple consequences of Eq. \eqref{eq:Sshift}: for $n>0$ we have
\begin{align}
\begin{pmatrix}
\mathcal{Q}_S^{\{n\}\left[1\right]}\\
\mathcal{Q}_S^{\{n\}\left[-1\right]}
\end{pmatrix} 
& =\begin{pmatrix}S+n-iu & -iu\\
-iu & -S-n-iu
\end{pmatrix}
\begin{pmatrix}
\mathcal{Q}_S^{\{n-1\}\left[1\right]}\\
\mathcal{Q}_S^{\{n-1\}\left[-1\right]}
\end{pmatrix}\,,\\
\begin{pmatrix}
\mathcal{Q}_S^{\{-n\}\left[1\right]}\\
\mathcal{Q}_S^{\{-n\}\left[-1\right]}
\end{pmatrix} 
& =\begin{pmatrix}S-n+1+iu & -iu\\
-iu & -S+n-1+iu
\end{pmatrix}
\begin{pmatrix}
\mathcal{Q}_S^{\{-n+1\}\left[1\right]}\\
\mathcal{Q}_S^{\{-n+1\}\left[-1\right]}
\end{pmatrix}\,.
\end{align}
These relations allow one to shift the upper index of $\mathcal{Q}_S^{\{n\}}$ to zero without generating any denominators. Since  $\mathcal{Q}_S^{\{0\}}=Q_S$, we use these identities for the reduction of sums \eqref{eq:calQ1} \eqref{eq:calQ1} to the combination of $\left\langle {Q}|\ldots\right\rangle$, possibly, with shifted argument and/or multiplied by powers of $u$.

The sums of the form \eqref{eq:calQ2} after the substitution of the definition \eqref{eq:calQdef} and shifting $j_1\to j_1+n$ are almost of the required form except for the  upper limit of summation over $j_2$ which is $j_1+n-1$, i.e., is shifted by $n$. Then we can treat the missing/redundant terms in a recursive manner. For example, we have
\begin{multline}
	\left\langle \mathcal{Q}^{\left\{-1\right\} }|\tilde{1}\left(\j+1\right),w_1(\j),W\right\rangle=
	\left\langle Q|(S+1-\j)\tilde{1}\left(\j\right),w_1(\j),W\right\rangle\\
	-\left\langle \mathcal{Q}^{\left\{-1\right\} }|\tilde{1}\left(\j+1\right)w_1(\j),W\right\rangle\,.
\end{multline}
Then the second term is again of the same form as those in the right-hand side of Eq. \eqref{eq:calQinrhs}. Note that special attention should be paid to the sums with depth less or equal to $|n|$.
%TODO check/improve
The full set of the reductions rules can be found in the code of the attached {\it Mathematica} file.

\subsection{Constraints solution}
\label{constraintsSolution}

Now, with the knowledge of how to find the solutions of two Baxter equations \eqref{Baxter1a} and \eqref{Baxter2a} in each order of perturbation theory we may proceed with the determination of constants in the anzats for $\bP$ - functions together with additional $\phi_{a,j}^{\mathrm{per|anti}}$ constants parameterizing homogeneous parts of the solutions to mentioned Baxter equations. It should be noted that to solve constraint equations one can greatly benefit from the use of elementary operations on Baxter polynomials and their sums, such as shifts and partial fractions, see subsection \ref{elementaryopSection}. In addition we have also extensively used stuffle relations for products of $\langle w,W\rangle$ - sums together with Taylor expansion of 
$\langle Q|w,W\rangle$ sums at $u=i/2$, see corresponding derivative rules at the end of subsection \ref{elementaryopSection}. To have intermediate expressions as small as possible we have also fixed residual gauge freedom by putting all coefficients $m_j^{(l)}$, $k_j^{(l)}$ to zero.

In the beginning of subsection \ref{elementaryopSection} we have already found the expression for $A_0^{(0)}$ constant together with LO solution of the first Baxter equation \eqref{Baxternu1} up to yet undetermined constant $\alpha$. Next, from equation (\ref{nu3sol}) we determine the expression for $\nu_3^{(0)} (u)$ and substitute it in the equation (\ref{pt2}). Expanding the latter at $u=0$ up to $\OO (u^2)$ we get the expression\footnote{See appendix \ref{H-B-sums} for definition of $H$ and $B$ sums.}  for constant $\alpha$:
\begin{equation}
\frac{1}{\alpha^2} = -4 i B_1\, ,\quad B_1 = H_1 - H_{-1}\, . \label{alpha}
\end{equation}
Also from the requirement of absence of poles in combinations (\ref{cutsfree}) for $\nu_1^{(0)}$ we may determine the value of $\s = (-)^S$ in Eq. \eqref{nuanalytcont}.  Knowing the expression for $\nu_1^{(0)} (u)$ we may determine $\nu_2^{(0)}$ by solving second Baxter equation (\ref{Baxternu2}): 
\begin{multline}
\nu_{2}^{\left(0\right)}  =-(-)^S\frac{\alpha}{8}A_3^{(0)}\bigg[\frac{3\left(S+2\right)\left(S+1\right)}{2S+3}\,Q_{S+2}^{[-1]}-2\frac{3S^{2}+3S+1}{2S+1}Q_S^{[-1]}  +\frac{3S\,\left(S-1\right)}{2S-1}Q_{S-2}^{[-1]}\bigg]
\end{multline}
Now, from equation (\ref{pt1}) expanded at  $u=0$ up to $\OO (u)$ we get the value of $A_3^{(0)}$ constant:
\begin{equation}
A_3^{(0)} = -\frac{4}{3} (2S+3)(2S-1) B_1 
\end{equation} 

Starting from NLO, the solution at each perturbation order follows the same pattern. First, the required expressions for $\nu_3^{(k)}$ and $\nu_4^{(k)}$ functions at each perturbation order are obtained from the equations \eqref{nu3sol} and \eqref{nu4sol}, respectively,expanded to the required order in QSC coupling constant. Then the required steps at each perturbation order can be summarized as follows:

\begin{enumerate}
	\item  To get expressions for additional coefficients $c_{2,n}^{(l)}$, $l<i$ required at a given perturbation order $k$ we perform a number of expansions of equation \eqref{pt2} up to $\OO(h^{1+2 l})$ and $\OO (u^{2+2(k-l)})$ for all $l<k$. 
	\item To get expressions for additional coefficients $c_{1,n}^{(l)}$, $l<k$ required at a given perturbation order $k$ we perform a number of expansions of equation \eqref{pt1} up to $\OO(h^{1+2 l})$ and $\OO (u^{1+2(k-l)})$ for all $l<k$. 
	\item To get expressions for additional coefficients $c_{0,n}^{(l)}$, $l<k$ required at a given perturbation order $k$ we perform an expansion of equation \eqref{p0} up to $\OO (h^{1+2k})$ and $\OO (u)$. Here, we would like to note, that combinations in brackets of equation \eqref{p0} are the same as in equations \eqref{cutsfree} and therefore they are free of poles at $u=0$ at any order of perturbation theory. The leading order $\OO (h^0)$ expressions for $\bp$'s and $hx$ are also regular at $u=0$. So, up to $\OO (u)$ order of expansion in $u$ the $\nu_a^{(k)}$ - functions do not contribute to the resulting equations on coefficients $c_{0,n}^{(l)}$.
	\item To get expressions for additional coefficients $c_{3,n}^{(l)}$, $l<k$ required at a given perturbation order $k$ we perform an expansion of equation \eqref{p3} up to $\OO (h^{1+2k})$ and $\OO (u)$. Here apply the same argument we did at the end of the previous step and  up to $\OO (u)$ order of expansion in $u$ the $\nu_a^{(k)}$ - functions do not contribute to the resulting equations on coefficients $c_{3,n}^{(l)}$.
	\item Requiring the absence of poles in the combinations \eqref{cutsfree} for $\nu_1^{(k)}$ function allows us to fix $A_0^{(k)}$ together with expressions for all constants in the homogeneous piece of solution for $\nu_1^{(k)}$ except the value of $\phi_{1,0}^{(k),\mathrm{per}}$.
	\item To determine the value of $\phi_{1,0}^{(k),\mathrm{per}}$ coefficient we perform the expansion of equation \eqref{pt2} up to the order $\OO (h^{1+2k})$ and $\OO (u^2)$. 
	\item Requiring the absence of poles in the combinations \eqref{cutsfree} for $\nu_2^{(k)}$ function allows us to fix expressions for all constants in the homogeneous piece of solution for $\nu_2^{(k)}$.
	\item Finally, to get the expression for $A_3^{(k)}$ coefficient we perform the expansion of equation \eqref{pt1} up to the order $\OO (h^{1+2k})$ and $\OO (u)$. 
\end{enumerate}

\subsubsection{NLO}

Following the above procedure step by step at NLO  we get\footnote{The expressions for $q_{1,2}^{(1|2)}$ - functions can be found in accompanying {\it Mathematica} files.}

\begin{enumerate}
	\item First, from equation \eqref{pt2} :
	\begin{equation}
	c_{2,2}^{(0)} = 0\, ,\quad c_{2,3}^{(0)} = 4 i\alpha^2 \left(
	B_1  B_2  + B_3 
	\right) \, .
	\end{equation}
	\item Then, from equation \eqref{pt1} :
	\begin{equation}
	c_{1,1}^{(0)} = 0\, ,\quad
	c_{1,2}^{(0)} = 4B_1-B_1^2-2B_2
	%4 H_{-1} (S) - H_{-1} (S)^2 + 4 H_1 (S) + 2 H_{-1} (S) H_{1} (S) - H_{1} (S)^2 + 2 H_{-1,-1} (S) - 2 H_{-1,1} (S) + 2 H_{1,-1} (S) - 2 H_{1,1} (S)
	\, .
	\end{equation}
	\item Equation \eqref{p0} gives
	\begin{equation}
	c_{0,1}^{(0)} = - 1 - \frac{i (-)^S S\, (1+S)}{3 (1+ 2 S) \alpha^2}\, .
	\end{equation}
	\item  Similarly from equation \eqref{p3} we have 
	\begin{equation}
	c_{3,1}^{(0)} = \frac{2 (-)^S S (1+S) (12 i (1 + 2 S)\alpha^2 - 2 (-)^S S (1+S) )}{36 (1+ 2 S)^2\alpha^4}\, .
	\end{equation}
	\item The homogeneous part of $q_1^{(1)} = \left(\nu_1^{(1)}\right)^{[1]}$ function is given by 
	\begin{equation}
	q_{1,hom}^{(1)} = \alpha\phi_{1,0}^{(1),\mathrm{per}} Q_S + \alpha \phi_{1,1}^{(1),\mathrm{per}}\P_1 (u+i/2) Q_S 
	\end{equation}
	and from the absence of poles in equations \eqref{cutsfree} we have 
	\begin{equation}
	A_0^{(1)} = 2 (-)^S\, (3 + 2 S) B_1\, ,\quad \phi_{1,1}^{(1), \mathrm{per}} = 2 i B_1\, .
	\end{equation}
	\item From equation \eqref{pt2} we further fix\footnote{This expression is different from the one reported in \cite{ABJM_QSC_Mellin} due to different particular solution contribution.} 
	\begin{multline}
	\phi_{1,0}^{(1),\mathrm{per}} = \frac{1}{B_1}\Bigg\{
	-8 H_{-2,1}+2 H_{-1,-2}-4 H_{-1,-1}+4 H_{-1,1}-2 H_{-1,2}-6 H_{1,-2} \\  +4
	H_{1,-1}  -4 H_{1,1}+6 H_{1,2}+8 H_{2,1}-4 H_{-1,-1,-1}-4 H_{-1,-1,1}  +8
	H_{-1,1,1} \\  +8 H_{1,-1,1}  +4 H_{1,1,-1}-12 H_{1,1,1}+2 H_{-3}-4 H_{-2}+4 H_2-2
	H_3 \Bigg\}\, .
	\end{multline}
	\item The homogeneous part of $q_2^{(1)} = \left(\nu_2^{(1)}\right)^{[1]}$ function is given by 
	\begin{equation}
	q_{2,hom}^{(1)} = \alpha\phi_{2,0}^{(1),\mathrm{per}} \Z_S + \alpha \phi_{2,0}^{(1),\mathrm{anti}}\P_{-1} (u+i/2) Q_S 
	\end{equation}
	and from the absence of poles in equations \eqref{cutsfree} we have 
	\begin{equation}
	\phi_{2,0}^{(1),\mathrm{per}} = 0\, , \quad \phi_{2,0}^{(1),\mathrm{anti}} = 4 i B_1\, .
	\end{equation}
	\item Finally, from equation \eqref{pt1} we have
	\begin{multline}
	A_3^{(1)} = 
	-\frac{16}{3} (2 S-1) (2 S+3)\Big(3 \bar{H}_{-2,-1}-2
	\bar{H}_{-2,i}-\bar{H}_{-2,1}-\bar{H}_{-1,-2
	}+2 \bar{H}_{-1,2 i}-\bar{H}_{-1,2}
	\\
	-6
	\bar{H}_{i,-2}+12 \bar{H}_{i,2 i}-6
	\bar{H}_{i,2}-6 \bar{H}_{2 i,-1}+4
	\bar{H}_{2 i,i}+2 \bar{H}_{2
		i,1}-\bar{H}_{1,-2}+2 \bar{H}_{1,2
		i}-\bar{H}_{1,2}+3 \bar{H}_{2,-1}
	\\
	-2
	\bar{H}_{2,i}-\bar{H}_{2,1}+2
	\bar{H}_{-1,i,-1}-2 \bar{H}_{-1,i,1}+8
	\bar{H}_{i,-1,-1}-12 \bar{H}_{i,-1,i}+4
	\bar{H}_{i,-1,1}-16 \bar{H}_{i,i,-1}\\
	+16
	\bar{H}_{i,i,i}+4 \bar{H}_{i,1,-1}-4
	\bar{H}_{i,1,i}+2 \bar{H}_{1,i,-1}-2
	\bar{H}_{1,i,1}
	-\frac12 B_1 \zeta_2
	\Big)- \frac{4}{3} (5 + 20 S + 4 S^2) B_1^2
	\end{multline}
\end{enumerate}

\subsubsection{NNLO}

At NNLO we were not reducing all $\left\langle w_{1}\left(\j\right),w_{2}\left(\j\right),\ldots,w_{n}\left(\j\right)\right\rangle$ sums at intermediate steps to $H$ and $B$ -sums. Such reduction was performed only for the NNLO anomalous dimensions at the end. Moreover, this final reduction was not algorithmic - we just solved a system of equations for $768$ spin values, which is the dimension of our $\barH$ basis\footnote{See definition of $\barH$ - sums in appendix \ref{H-B-sums}.} at weight 5 corresponding to NNLO. Still, our preliminary considerations show that the required algorithmic reduction at all steps is possible and, what is more important, it will make our algorithm much more efficient. The details of this reduction will be the subject of one of our subsequent papers. Following the steps of general procedure for constraints solution at NNLO we get\footnote{The files $\text{c}ijn$ contain results for $c_{i,j}^{(n)}$ coefficients, $\text{A}in$ for $A_i^{(n)}$, $\text{phiP}ani$ for $\phi_{a,i}^{(n),\mathrm{per}}$ and $\text{phiA}ani$ for $\phi_{a,i}^{(n),\mathrm{anti}}$.}
\begin{enumerate}
	\item First, from equation \eqref{pt2}:
	\begin{equation}
	c_{2,4}^{(0)} = 0\, , \quad c_{2,2}^{(1)} = 0
	\end{equation} 
	and the results for coefficients $c_{2,5}^{(0)}$ and $c_{2,3}^{(1)}$ can be found in accompanying {\it Mathematica} files.
	\item
	Then, from equation \eqref{pt1}:
	\begin{equation}
	c_{1,3}^{(0)} = 0\, , \quad c_{1,1}^{(1)} = 0
	\end{equation} 
	and the results for coefficients $c_{1,4}^{(0)}$ and $c_{1,2}^{(1)}$ can be found in accompanying {\it Mathematica} files.
	\item Next, equation \eqref{p0} gives $c_{0,2}^{(0)}=0$ and the expression for $c_{0,1}^{(1)}$ may be found in accompanying {\it Mathematica} file.  
	\item Similarly equation \eqref{p3} gives $c_{3,2}^{(0)} = 0$ and the expression for $c_{3,1}^{(1)}$ may be found in accompanying {\it Mathematica} file. 
	\item  The homogeneous part of $q_1^{(2)} = \left(\nu_1^{(2)}\right)^{[1]}$ function is given by 
	\begin{equation}
	q_{1,hom}^{(2)} = \alpha\phi_{1,0}^{(1),\mathrm{per}} Q_S + \alpha\sum_{i=1}^3 \phi_{1,i}^{(2),\mathrm{per}}\P_i (u+i/2) Q_S 
	+ \alpha\sum_{i=1}^2 \phi_{1,i-1}^{(2),\mathrm{anti}}\P_{-i} (u+i/2)\Z_S
	\end{equation}
	and from the absence of poles in equations \eqref{cutsfree} we have 
	\begin{equation}
    \phi_{1,3}^{(2), \mathrm{per}} = 2 i B_1\, ,\quad \phi_{1,0}^{(2), \mathrm{anti}} = 0\, ,\quad \phi_{1,1}^{(2), \mathrm{anti}} = - 4 i B_1\,  
	\end{equation}
	while the expressions for $A_0^{(2)}$,   $\phi_{1,1}^{(2), \mathrm{per}}$ and $\phi_{1,2}^{(2), \mathrm{per}}$ can be found in accompanying {\it Mathematica} files. 
	\item From equation \eqref{pt2} we further fix the expression for $\phi_{1,0}^{(2), \mathrm{per}}$, which  may be found in accompanying {\it Mathematica} file. 
	\item  The homogeneous part of $q_2^{(2)} = \left(\nu_2^{(2)}\right)^{[1]}$ function is given by 
	\begin{equation}
	q_{2,hom}^{(2)} = \alpha\phi_{2,0}^{(2),\mathrm{per}} \Z_S + \alpha\sum_{i=1}^2\phi_{2,i}^{(2),\mathrm{per}}\P_i (u+i/2)\Z_S 
	+ \alpha\sum_{i=1}^3\phi_{2,i-1}^{(2),\mathrm{anti}}\P_{-i} (u+i/2) Q_S 
	\end{equation}
	and from the absence of poles in equations \eqref{cutsfree} we have 
	\begin{equation}
	\phi_{2,0}^{(2),\mathrm{per}} = \phi_{2,2}^{(2),\mathrm{per}} = \phi_{2,1}^{(2),\mathrm{anti}}= 0\, , \quad \phi_{2,2}^{(2),\mathrm{anti}}= 4 i B_1\, ,
	\end{equation}
	while the expressions for $\phi_{2,1}^{(2),\mathrm{per}}$ and $\phi_{2,0}^{(2),\mathrm{anti}}$ can be found in accompanying {\it Mathematica} files. 
	\item  Finally, from equation \eqref{pt1} we determine the value of $A_3^{(2)}$, which again may be found in accompanying {\it Mathematica} file. 
\end{enumerate}

\section{Anomalous dimensions}\label{AnomalousDimensions}
\noindent
The expressions for the anomalous dimensions can be easily obtained from the corresponding expressions for $A_3^{(0,1,2)}$ constants with the help of equation (\ref{Pasympt}). The results for anomalous dimensions of twist 1 operators up to six loop are then given by\footnote{See appendix \ref{H-B-sums} for the definition of $\bar H$ - sums.}
\begin{equation}
\gamma (S) = \gamma^{(0)} (S) h^2 + \gamma^{(1)} (S) h^4 + \gamma^{(2)} (S) h^6 +\ldots
\end{equation}
where
\begin{equation}
\gamma^{(0)}(S) = 4 \left(\barH_1 + \barH_{-1} - 2\barH_i
\right)
\end{equation}
\begin{multline}
\gamma^{(1)}(S) = 16 \Big\{ 3 \barH_{-2,-1} - 2 \barH_{-2, i} - \barH_{-2,1} - \barH_{-1,-2} + 2 \barH_{-1, 2 i} - \barH_{-1,2} - 6 \barH_{i,-2} \\
+ 12 \barH_{i, 2 i} - 6 \barH_{i, 2} - 6 \barH_{2 i, -1} + 4 \barH_{2 i, i} + 2 \barH_{2 i, 1} - \barH_{1,-2} + 2 \barH_{1, 2 i} - \barH_{1,2} + 3\barH_{2,-1} \\ - 2 \barH_{2, i} - \barH_{2,1} + 2\barH_{-1, i, -1} - 2 \barH_{-1, i, 1} + 8\barH_{i, -1, -1} - 12\barH_{i, -1, i} + 4 \barH_{i, -1, 1} - 16\barH_{i, i, -1} \\
+ 16 \barH_{i, i, i} + 4 \barH_{i, 1, -1} - 4 \barH_{i, 1, i} + 2 \barH_{1,i,-1} - 2 \barH_{1,i,1} \Big\} - \frac{4}{3}\pi^2 \left(
\barH_{-1} + \barH_{1} - 2\barH_{i}
\right) \\
\end{multline}
and
\begin{equation}
\gamma^{(2)}(S) = \{1\} + \{2\} + \{3\} + \{4\} + \{5\}\, 
\end{equation}
with 
\begin{align}
\{1\} = \frac{7}{5} \pi ^4 \left(\bar{H}_1+\bar{H}_{-1}-2 \bar{H}_i\right)\, ,
\end{align}
\begin{multline}
\{2\} = -144 \zeta (3) \Big\{-\bar{H}_{-1,-1}+2 \bar{H}_{-1,i}-\bar{H}_{-1,1}+2
\bar{H}_{i,-1}-4 \bar{H}_{i,i}+2 \bar{H}_{i,1} \\ -\bar{H}_{1,-1} +2
\bar{H}_{1,i}-\bar{H}_{1,1}+2 \bar{H}_{-2}-4 \bar{H}_{2 i}+2 \bar{H}_2\Big\}\, ,
\end{multline}
\begin{multline}
\{3\} = 32 \pi ^2 \log (2) \Big\{-\bar{H}_{-1,-1}+2 \bar{H}_{-1,i}-\bar{H}_{-1,1}+2
\bar{H}_{i,-1}-4 \bar{H}_{i,i}+2 \bar{H}_{i,1} \\ -\bar{H}_{1,-1}+2
\bar{H}_{1,i}-\bar{H}_{1,1}+2 \bar{H}_{-2}-4 \bar{H}_{2 i}+2 \bar{H}_2\Big\}\, ,
\end{multline}
\begin{multline}
\{4\} = \frac{16}{3} \pi ^2 \Big\{-16 \bar{H}_{-2,-1}+48 \bar{H}_{-2,i}-8 \bar{H}_{-2,1}-6
\bar{H}_{-1,-2}+12 \bar{H}_{-1,2 i}-6 \bar{H}_{-1,2} \\ +60 \bar{H}_{i,-2} -120
\bar{H}_{i,2 i}+60 \bar{H}_{i,2}+32 \bar{H}_{2 i,-1}-96 \bar{H}_{2 i,i}+16
\bar{H}_{2 i,1}-6 \bar{H}_{1,-2}\\ +12 \bar{H}_{1,2 i}  -6 \bar{H}_{1,2} -16
\bar{H}_{2,-1}+48 \bar{H}_{2,i}-8 \bar{H}_{2,1}-3 \bar{H}_{-1,-1,-1}-6
\bar{H}_{-1,-1,i} \\ -3 \bar{H}_{-1,-1,1}-14 \bar{H}_{-1,i,-1}+44 \bar{H}_{-1,i,i}-6
\bar{H}_{-1,i,1}-3 \bar{H}_{-1,1,-1}-6 \bar{H}_{-1,1,i}\\ -3 \bar{H}_{-1,1,1}-26
\bar{H}_{i,-1,-1}+68 \bar{H}_{i,-1,i}-18 \bar{H}_{i,-1,1}+84 \bar{H}_{i,i,-1}-184
\bar{H}_{i,i,i}\\+52 \bar{H}_{i,i,1}-18 \bar{H}_{i,1,-1}+52 \bar{H}_{i,1,i}-10
\bar{H}_{i,1,1}-3 \bar{H}_{1,-1,-1}-6 \bar{H}_{1,-1,i}\\-3 \bar{H}_{1,-1,1}-14
\bar{H}_{1,i,-1}+44 \bar{H}_{1,i,i}-6 \bar{H}_{1,i,1}-3 \bar{H}_{1,1,-1}-6
\bar{H}_{1,1,i}-3 \bar{H}_{1,1,1}\\+20 \bar{H}_{-3}-40 \bar{H}_{3 i}+20
\bar{H}_3\Big\}\, ,
\end{multline}
and
\begin{multline}
\{5\} = 64 \Big\{12 \bar{H}_{-3,-2}+8 \bar{H}_{-3,2 i}-20 \bar{H}_{-3,2}-12
\bar{H}_{-2,-3}+24 \bar{H}_{-2,3 i}-12 \bar{H}_{-2,3}+4 \bar{H}_{-1,-4}\\  -8
\bar{H}_{-1,4 i}+4 \bar{H}_{-1,4}+8 \bar{H}_{i,-4}-16 \bar{H}_{i,4 i}+8
\bar{H}_{i,4}-8 \bar{H}_{2 i,-3}+16 \bar{H}_{2 i,3 i}-8 \bar{H}_{2 i,3}\\  -24
\bar{H}_{3 i,-2}-16 \bar{H}_{3 i,2 i}+40 \bar{H}_{3 i,2}+4 \bar{H}_{1,-4}-8
\bar{H}_{1,4 i}+4 \bar{H}_{1,4}-12 \bar{H}_{2,-3}+24 \bar{H}_{2,3 i}\\  -12
\bar{H}_{2,3}+12 \bar{H}_{3,-2}+8 \bar{H}_{3,2 i}-20 \bar{H}_{3,2}+18
\bar{H}_{-3,-1,-1}-12 \bar{H}_{-3,-1,i}-6 \bar{H}_{-3,-1,1} \\  -28
\bar{H}_{-3,i,-1}  +24 \bar{H}_{-3,i,i}+4 \bar{H}_{-3,i,1}-6 \bar{H}_{-3,1,-1}+4
\bar{H}_{-3,1,i}+2 \bar{H}_{-3,1,1}+30 \bar{H}_{-2,-2,-1} \\  -24 \bar{H}_{-2,-2,i}-6
\bar{H}_{-2,-2,1}+22 \bar{H}_{-2,-1,-2}-60 \bar{H}_{-2,-1,2 i}+38
\bar{H}_{-2,-1,2}+4 \bar{H}_{-2,i,-2} \\  +56 \bar{H}_{-2,i,2 i}  -60
\bar{H}_{-2,i,2}-44 \bar{H}_{-2,2 i,-1}+48 \bar{H}_{-2,2 i,i}-4 \bar{H}_{-2,2
	i,1}-2 \bar{H}_{-2,1,-2}  \\  -12 \bar{H}_{-2,1,2 i}  +14 \bar{H}_{-2,1,2}+46
\bar{H}_{-2,2,-1}-56 \bar{H}_{-2,2,i}+10 \bar{H}_{-2,2,1}-12
\bar{H}_{-1,-3,-1} \\  +12 \bar{H}_{-1,-3,i}-15 \bar{H}_{-1,-2,-2}+14 \bar{H}_{-1,-2,2
	i}+\bar{H}_{-1,-2,2}-4 \bar{H}_{-1,-1,-3}+8 \bar{H}_{-1,-1,3 i}\\  -4
\bar{H}_{-1,-1,3}+16 \bar{H}_{-1,i,-3}-32 \bar{H}_{-1,i,3 i}+16
\bar{H}_{-1,i,3}+34 \bar{H}_{-1,2 i,-2}-36 \bar{H}_{-1,2 i,2 i}\\  +2 \bar{H}_{-1,2
	i,2}+24 \bar{H}_{-1,3 i,-1}-24 \bar{H}_{-1,3 i,i}-4 \bar{H}_{-1,1,-3}+8
\bar{H}_{-1,1,3 i}-4 \bar{H}_{-1,1,3}\\  -15 \bar{H}_{-1,2,-2}+14 \bar{H}_{-1,2,2
	i}+\bar{H}_{-1,2,2}-12 \bar{H}_{-1,3,-1}+12 \bar{H}_{-1,3,i}-64
\bar{H}_{i,-3,-1}\\ +56 \bar{H}_{i,-3,i}+8 \bar{H}_{i,-3,1}-10 \bar{H}_{i,-2,-2}+84
\bar{H}_{i,-2,2 i}-74 \bar{H}_{i,-2,2}-56 \bar{H}_{i,-1,-3} \\ +112 \bar{H}_{i,-1,3
	i}-56 \bar{H}_{i,-1,3}+48 \bar{H}_{i,i,-3}-96 \bar{H}_{i,i,3 i}+48
\bar{H}_{i,i,3}+44 \bar{H}_{i,2 i,-2}-216 \bar{H}_{i,2 i,2 i}\\ +172 \bar{H}_{i,2
	i,2}+128 \bar{H}_{i,3 i,-1}-112 \bar{H}_{i,3 i,i}-16 \bar{H}_{i,3 i,1}-8
\bar{H}_{i,1,-3}+16 \bar{H}_{i,1,3 i}  -8 \bar{H}_{i,1,3} \\ -42 \bar{H}_{i,2,-2}+148
\bar{H}_{i,2,2 i}-106 \bar{H}_{i,2,2}-64 \bar{H}_{i,3,-1}+56 \bar{H}_{i,3,i}+8
\bar{H}_{i,3,1}-28 \bar{H}_{2 i,-2,-1} \\ +28 \bar{H}_{2 i,-2,1}-12 \bar{H}_{2
	i,-1,-2}+56 \bar{H}_{2 i,-1,2 i}-44 \bar{H}_{2 i,-1,2}-56 \bar{H}_{2 i,i,-2}-16
\bar{H}_{2 i,i,2 i}\\ +72 \bar{H}_{2 i,i,2}  +24 \bar{H}_{2 i,2 i,-1}-24 \bar{H}_{2
	i,2 i,1}+20 \bar{H}_{2 i,1,-2}-8 \bar{H}_{2 i,1,2 i}-12 \bar{H}_{2 i,1,2}  -60
\bar{H}_{2 i,2,-1} \\ +64 \bar{H}_{2 i,2,i}-4 \bar{H}_{2 i,2,1}-36 \bar{H}_{3
	i,-1,-1}+24 \bar{H}_{3 i,-1,i}  +12 \bar{H}_{3 i,-1,1}+56 \bar{H}_{3 i,i,-1}-48
\bar{H}_{3 i,i,i} \\ -8 \bar{H}_{3 i,i,1}+12 \bar{H}_{3 i,1,-1}-8 \bar{H}_{3 i,1,i}-4
\bar{H}_{3 i,1,1}-12 \bar{H}_{1,-3,-1}+12 \bar{H}_{1,-3,i}-15
\bar{H}_{1,-2,-2} \\ +14 \bar{H}_{1,-2,2 i}+\bar{H}_{1,-2,2}-4 \bar{H}_{1,-1,-3}+8
\bar{H}_{1,-1,3 i}-4 \bar{H}_{1,-1,3}+16 \bar{H}_{1,i,-3}-32 \bar{H}_{1,i,3 i} \\ +16
\bar{H}_{1,i,3}+34 \bar{H}_{1,2 i,-2}-36 \bar{H}_{1,2 i,2 i}+2 \bar{H}_{1,2
	i,2}+24 \bar{H}_{1,3 i,-1}-24 \bar{H}_{1,3 i,i}-4 \bar{H}_{1,1,-3} \\ +8
\bar{H}_{1,1,3 i}-4 \bar{H}_{1,1,3}-15 \bar{H}_{1,2,-2}+14 \bar{H}_{1,2,2
	i}+\bar{H}_{1,2,2}-12 \bar{H}_{1,3,-1}+12 \bar{H}_{1,3,i} \\ +30 \bar{H}_{2,-2,-1}-24
\bar{H}_{2,-2,i}-6 \bar{H}_{2,-2,1}+22 \bar{H}_{2,-1,-2}-60 \bar{H}_{2,-1,2 i}+38
\bar{H}_{2,-1,2}+4 \bar{H}_{2,i,-2} \\ +56 \bar{H}_{2,i,2 i}-60 \bar{H}_{2,i,2}-44
\bar{H}_{2,2 i,-1}+48 \bar{H}_{2,2 i,i}-4 \bar{H}_{2,2 i,1}-2 \bar{H}_{2,1,-2}-12
\bar{H}_{2,1,2 i} \\ +14 \bar{H}_{2,1,2}+46 \bar{H}_{2,2,-1}-56 \bar{H}_{2,2,i}+10
\bar{H}_{2,2,1}+18 \bar{H}_{3,-1,-1}-12 \bar{H}_{3,-1,i}-6 \bar{H}_{3,-1,1} \\ -28
\bar{H}_{3,i,-1}+24 \bar{H}_{3,i,i}+4 \bar{H}_{3,i,1}-6 \bar{H}_{3,1,-1}+4
\bar{H}_{3,1,i}+2 \bar{H}_{3,1,1}-48 \bar{H}_{-2,-1,-1,-1} \\ +64
\bar{H}_{-2,-1,-1,i}-16 \bar{H}_{-2,-1,-1,1}+80 \bar{H}_{-2,-1,i,-1}-80
\bar{H}_{-2,-1,i,i}-16 \bar{H}_{-2,-1,1,-1}+16 \bar{H}_{-2,-1,1,i} \\ +52
\bar{H}_{-2,i,-1,-1}-48 \bar{H}_{-2,i,-1,i}-4 \bar{H}_{-2,i,-1,1}-112
\bar{H}_{-2,i,i,-1}+112 \bar{H}_{-2,i,i,i}-4 \bar{H}_{-2,i,1,-1}\\ +4
\bar{H}_{-2,i,1,1}-16 \bar{H}_{-2,1,-1,-1}+16 \bar{H}_{-2,1,-1,i}+16
\bar{H}_{-2,1,i,-1}-16 \bar{H}_{-2,1,i,i}-8 \bar{H}_{-1,-2,-1,i} \\ +8
\bar{H}_{-1,-2,-1,1}-10 \bar{H}_{-1,-2,i,-1}+12 \bar{H}_{-1,-2,i,i}-2
\bar{H}_{-1,-2,i,1}+8 \bar{H}_{-1,-2,1,-1}-8 \bar{H}_{-1,-2,1,i} \\ +4
\bar{H}_{-1,-1,-2,-1}-4 \bar{H}_{-1,-1,-2,i}+3 \bar{H}_{-1,-1,-1,-2}-6
\bar{H}_{-1,-1,-1,2 i}+3 \bar{H}_{-1,-1,-1,2}-18 \bar{H}_{-1,-1,i,-2}\\ +20
\bar{H}_{-1,-1,i,2 i}-2 \bar{H}_{-1,-1,i,2}-24 \bar{H}_{-1,-1,2 i,-1}+24
\bar{H}_{-1,-1,2 i,i}+3 \bar{H}_{-1,-1,1,-2}-6 \bar{H}_{-1,-1,1,2 i}\\ +3
\bar{H}_{-1,-1,1,2}+4 \bar{H}_{-1,-1,2,-1}-4 \bar{H}_{-1,-1,2,i}-6
\bar{H}_{-1,i,-2,-1}+24 \bar{H}_{-1,i,-2,i}-18 \bar{H}_{-1,i,-2,1}\\ -8
\bar{H}_{-1,i,-1,-2}+8 \bar{H}_{-1,i,-1,2}+56 \bar{H}_{-1,i,i,-2}-48
\bar{H}_{-1,i,i,2 i}-8 \bar{H}_{-1,i,i,2}+28 \bar{H}_{-1,i,2 i,-1}\\ -48
\bar{H}_{-1,i,2 i,i}+20 \bar{H}_{-1,i,2 i,1}-16 \bar{H}_{-1,i,1,-2}+16
\bar{H}_{-1,i,1,2 i}+10 \bar{H}_{-1,i,2,-1}-8 \bar{H}_{-1,i,2,i}\\ -2
\bar{H}_{-1,i,2,1}-6 \bar{H}_{-1,2 i,-1,-1}+24 \bar{H}_{-1,2 i,-1,i}-18
\bar{H}_{-1,2 i,-1,1}+28 \bar{H}_{-1,2 i,i,-1}-32 \bar{H}_{-1,2 i,i,i}\\ +4
\bar{H}_{-1,2 i,i,1}-18 \bar{H}_{-1,2 i,1,-1}+16 \bar{H}_{-1,2 i,1,i}+2
\bar{H}_{-1,2 i,1,1}+4 \bar{H}_{-1,1,-2,-1}-4 \bar{H}_{-1,1,-2,i} \\ +3
\bar{H}_{-1,1,-1,-2}-6 \bar{H}_{-1,1,-1,2 i}+3 \bar{H}_{-1,1,-1,2}-18
\bar{H}_{-1,1,i,-2}+20 \bar{H}_{-1,1,i,2 i}-2 \bar{H}_{-1,1,i,2} \\ -24
\bar{H}_{-1,1,2 i,-1}+24 \bar{H}_{-1,1,2 i,i}+3 \bar{H}_{-1,1,1,-2}-6
\bar{H}_{-1,1,1,2 i}+3 \bar{H}_{-1,1,1,2}+4 \bar{H}_{-1,1,2,-1} \\ -4
\bar{H}_{-1,1,2,i}-8 \bar{H}_{-1,2,-1,i}+8 \bar{H}_{-1,2,-1,1}-10
\bar{H}_{-1,2,i,-1}+12 \bar{H}_{-1,2,i,i}-2 \bar{H}_{-1,2,i,1}  +8
\bar{H}_{-1,2,1,-1} \\ -8 \bar{H}_{-1,2,1,i}+94 \bar{H}_{i,-2,-1,-1}-104
\bar{H}_{i,-2,-1,i}+10 \bar{H}_{i,-2,-1,1}-188 \bar{H}_{i,-2,i,-1}+160
\bar{H}_{i,-2,i,i} \\ +28 \bar{H}_{i,-2,i,1}+10 \bar{H}_{i,-2,1,-1}-10
\bar{H}_{i,-2,1,1}+114 \bar{H}_{i,-1,-2,-1}-120 \bar{H}_{i,-1,-2,i}+6
\bar{H}_{i,-1,-2,1} \\ +76 \bar{H}_{i,-1,-1,-2}-168 \bar{H}_{i,-1,-1,2 i}+92
\bar{H}_{i,-1,-1,2}-112 \bar{H}_{i,-1,i,-2}+288 \bar{H}_{i,-1,i,2 i}-176
\bar{H}_{i,-1,i,2} \\ -212 \bar{H}_{i,-1,2 i,-1}+240 \bar{H}_{i,-1,2 i,i}-28
\bar{H}_{i,-1,2 i,1}+20 \bar{H}_{i,-1,1,-2}-56 \bar{H}_{i,-1,1,2 i}+36
\bar{H}_{i,-1,1,2} \\ +130 \bar{H}_{i,-1,2,-1}-152 \bar{H}_{i,-1,2,i}+22
\bar{H}_{i,-1,2,1}-152 \bar{H}_{i,i,-2,-1}+112 \bar{H}_{i,i,-2,i}+40
\bar{H}_{i,i,-2,1} \\ -92 \bar{H}_{i,i,-1,-2}+248 \bar{H}_{i,i,-1,2 i}-156
\bar{H}_{i,i,-1,2}+72 \bar{H}_{i,i,i,-2}-336 \bar{H}_{i,i,i,2 i}+264
\bar{H}_{i,i,i,2} \\ +304 \bar{H}_{i,i,2 i,-1}-288 \bar{H}_{i,i,2 i,i}-16
\bar{H}_{i,i,2 i,1}+4 \bar{H}_{i,i,1,-2}+56 \bar{H}_{i,i,1,2 i}-60
\bar{H}_{i,i,1,2}  -216 \bar{H}_{i,i,2,-1} \\ +240 \bar{H}_{i,i,2,i}-24
\bar{H}_{i,i,2,1}-192 \bar{H}_{i,2 i,-1,-1}+224 \bar{H}_{i,2 i,-1,i}-32
\bar{H}_{i,2 i,-1,1}+392 \bar{H}_{i,2 i,i,-1} \\ -368 \bar{H}_{i,2 i,i,i}-24
\bar{H}_{i,2 i,i,1}-32 \bar{H}_{i,2 i,1,-1}+32 \bar{H}_{i,2 i,1,i}+34
\bar{H}_{i,1,-2,-1}-24 \bar{H}_{i,1,-2,i}  -10 \bar{H}_{i,1,-2,1} \\ +20
\bar{H}_{i,1,-1,-2}-56 \bar{H}_{i,1,-1,2 i}+36 \bar{H}_{i,1,-1,2}+64
\bar{H}_{i,1,i,2 i}-64 \bar{H}_{i,1,i,2}-52 \bar{H}_{i,1,2 i,-1} +48
\bar{H}_{i,1,2 i,i} \\ +4 \bar{H}_{i,1,2 i,1}-4 \bar{H}_{i,1,1,-2}-8 \bar{H}_{i,1,1,2
	i}+12 \bar{H}_{i,1,1,2}+50 \bar{H}_{i,1,2,-1}-56 \bar{H}_{i,1,2,i}+6
\bar{H}_{i,1,2,1} \\ +110 \bar{H}_{i,2,-1,-1}-136 \bar{H}_{i,2,-1,i}+26
\bar{H}_{i,2,-1,1}-220 \bar{H}_{i,2,i,-1}+224 \bar{H}_{i,2,i,i}-4
\bar{H}_{i,2,i,1}+26 \bar{H}_{i,2,1,-1} \\ -32 \bar{H}_{i,2,1,i}+6
\bar{H}_{i,2,1,1}+72 \bar{H}_{2 i,-1,-1,-1}-88 \bar{H}_{2 i,-1,-1,i}+16
\bar{H}_{2 i,-1,-1,1}-120 \bar{H}_{2 i,-1,i,-1}+96 \bar{H}_{2 i,-1,i,i} \\ +24
\bar{H}_{2 i,-1,i,1}+16 \bar{H}_{2 i,-1,1,-1}-8 \bar{H}_{2 i,-1,1,i}-8 \bar{H}_{2
	i,-1,1,1}-64 \bar{H}_{2 i,i,-1,-1}+32 \bar{H}_{2 i,i,-1,i}+32 \bar{H}_{2
	i,i,-1,1} \\ +160 \bar{H}_{2 i,i,i,-1}-128 \bar{H}_{2 i,i,i,i}-32 \bar{H}_{2
	i,i,i,1}+32 \bar{H}_{2 i,i,1,-1}-32 \bar{H}_{2 i,i,1,i}+16 \bar{H}_{2
	i,1,-1,-1}-8 \bar{H}_{2 i,1,-1,i} \\ -8 \bar{H}_{2 i,1,-1,1}-8 \bar{H}_{2 i,1,i,-1}+8
\bar{H}_{2 i,1,i,1}-8 \bar{H}_{2 i,1,1,-1}+8 \bar{H}_{2 i,1,1,i}-8
\bar{H}_{1,-2,-1,i}+8 \bar{H}_{1,-2,-1,1} \\ -10 \bar{H}_{1,-2,i,-1}+12
\bar{H}_{1,-2,i,i}-2 \bar{H}_{1,-2,i,1}+8 \bar{H}_{1,-2,1,-1}-8
\bar{H}_{1,-2,1,i}+4 \bar{H}_{1,-1,-2,-1}-4 \bar{H}_{1,-1,-2,i} \\ +3
\bar{H}_{1,-1,-1,-2}-6 \bar{H}_{1,-1,-1,2 i}+3 \bar{H}_{1,-1,-1,2}-18
\bar{H}_{1,-1,i,-2}+20 \bar{H}_{1,-1,i,2 i}-2 \bar{H}_{1,-1,i,2}-24
\bar{H}_{1,-1,2 i,-1} \\ +24 \bar{H}_{1,-1,2 i,i}+3 \bar{H}_{1,-1,1,-2}-6
\bar{H}_{1,-1,1,2 i}+3 \bar{H}_{1,-1,1,2}+4 \bar{H}_{1,-1,2,-1}-4
\bar{H}_{1,-1,2,i}-6 \bar{H}_{1,i,-2,-1} \\ +24 \bar{H}_{1,i,-2,i}-18
\bar{H}_{1,i,-2,1}-8 \bar{H}_{1,i,-1,-2}+8 \bar{H}_{1,i,-1,2}+56
\bar{H}_{1,i,i,-2}-48 \bar{H}_{1,i,i,2 i}-8 \bar{H}_{1,i,i,2} \\ +28 \bar{H}_{1,i,2
	i,-1}-48 \bar{H}_{1,i,2 i,i}+20 \bar{H}_{1,i,2 i,1}-16 \bar{H}_{1,i,1,-2}+16
\bar{H}_{1,i,1,2 i}+10 \bar{H}_{1,i,2,-1}-8 \bar{H}_{1,i,2,i} \\ -2
\bar{H}_{1,i,2,1}-6 \bar{H}_{1,2 i,-1,-1}+24 \bar{H}_{1,2 i,-1,i}-18 \bar{H}_{1,2
	i,-1,1}+28 \bar{H}_{1,2 i,i,-1}-32 \bar{H}_{1,2 i,i,i}+4 \bar{H}_{1,2 i,i,1} \\ -18
\bar{H}_{1,2 i,1,-1}+16 \bar{H}_{1,2 i,1,i}+2 \bar{H}_{1,2 i,1,1}+4
\bar{H}_{1,1,-2,-1}-4 \bar{H}_{1,1,-2,i}+3 \bar{H}_{1,1,-1,-2}-6
\bar{H}_{1,1,-1,2 i} \\ +3 \bar{H}_{1,1,-1,2}-18 \bar{H}_{1,1,i,-2}+20
\bar{H}_{1,1,i,2 i}-2 \bar{H}_{1,1,i,2}-24 \bar{H}_{1,1,2 i,-1}+24 \bar{H}_{1,1,2
	i,i}+3 \bar{H}_{1,1,1,-2} \\ -6 \bar{H}_{1,1,1,2 i}+3 \bar{H}_{1,1,1,2}+4
\bar{H}_{1,1,2,-1}-4 \bar{H}_{1,1,2,i}-8 \bar{H}_{1,2,-1,i}+8
\bar{H}_{1,2,-1,1}-10 \bar{H}_{1,2,i,-1} \\ +12 \bar{H}_{1,2,i,i}-2
\bar{H}_{1,2,i,1}+8 \bar{H}_{1,2,1,-1}-8 \bar{H}_{1,2,1,i}-48
\bar{H}_{2,-1,-1,-1}+64 \bar{H}_{2,-1,-1,i}-16 \bar{H}_{2,-1,-1,1} \\ +80
\bar{H}_{2,-1,i,-1}-80 \bar{H}_{2,-1,i,i}-16 \bar{H}_{2,-1,1,-1}+16
\bar{H}_{2,-1,1,i}+52 \bar{H}_{2,i,-1,-1}-48 \bar{H}_{2,i,-1,i}-4
\bar{H}_{2,i,-1,1} \\ -112 \bar{H}_{2,i,i,-1}+112 \bar{H}_{2,i,i,i}-4
\bar{H}_{2,i,1,-1}+4 \bar{H}_{2,i,1,1}-16 \bar{H}_{2,1,-1,-1}+16
\bar{H}_{2,1,-1,i}+16 \bar{H}_{2,1,i,-1} \\ -16 \bar{H}_{2,1,i,i}+12
\bar{H}_{-1,-1,-1,i,-1}-12 \bar{H}_{-1,-1,-1,i,i}+6 \bar{H}_{-1,-1,i,-1,-1}-16
\bar{H}_{-1,-1,i,-1,i}+10 \bar{H}_{-1,-1,i,-1,1} \\ -16 \bar{H}_{-1,-1,i,i,-1}  +16
\bar{H}_{-1,-1,i,i,i}+10 \bar{H}_{-1,-1,i,1,-1}-8 \bar{H}_{-1,-1,i,1,i}-2
\bar{H}_{-1,-1,i,1,1}+12 \bar{H}_{-1,-1,1,i,-1} \\ -12 \bar{H}_{-1,-1,1,i,i}-24
\bar{H}_{-1,i,-1,-1,-1}+24 \bar{H}_{-1,i,-1,-1,i}+36 \bar{H}_{-1,i,-1,i,-1}-16
\bar{H}_{-1,i,-1,i,i}-20 \bar{H}_{-1,i,-1,i,1} \\ -8 \bar{H}_{-1,i,-1,1,i}+8
\bar{H}_{-1,i,-1,1,1}+32 \bar{H}_{-1,i,i,-1,i}-32 \bar{H}_{-1,i,i,-1,1}-24
\bar{H}_{-1,i,i,i,-1}+24 \bar{H}_{-1,i,i,i,1} \\ -32 \bar{H}_{-1,i,i,1,-1}+32
\bar{H}_{-1,i,i,1,i}-8 \bar{H}_{-1,i,1,-1,i}+8 \bar{H}_{-1,i,1,-1,1}-12
\bar{H}_{-1,i,1,i,-1}+16 \bar{H}_{-1,i,1,i,i} \\ -4 \bar{H}_{-1,i,1,i,1}+8
\bar{H}_{-1,i,1,1,-1}-8 \bar{H}_{-1,i,1,1,i}+12 \bar{H}_{-1,1,-1,i,-1}-12
\bar{H}_{-1,1,-1,i,i}+6 \bar{H}_{-1,1,i,-1,-1} \\ -16 \bar{H}_{-1,1,i,-1,i}+10
\bar{H}_{-1,1,i,-1,1}-16 \bar{H}_{-1,1,i,i,-1}+16 \bar{H}_{-1,1,i,i,i}+10
\bar{H}_{-1,1,i,1,-1}-8 \bar{H}_{-1,1,i,1,i} \\ -2 \bar{H}_{-1,1,i,1,1}+12
\bar{H}_{-1,1,1,i,-1}-12 \bar{H}_{-1,1,1,i,i}-96 \bar{H}_{i,-1,-1,-1,-1}+144
\bar{H}_{i,-1,-1,-1,i}-48 \bar{H}_{i,-1,-1,-1,1} \\ +180 \bar{H}_{i,-1,-1,i,-1}  -208
\bar{H}_{i,-1,-1,i,i}+28 \bar{H}_{i,-1,-1,i,1}-48 \bar{H}_{i,-1,-1,1,-1}+64
\bar{H}_{i,-1,-1,1,i}-16 \bar{H}_{i,-1,-1,1,1} \\ +192 \bar{H}_{i,-1,i,-1,-1}-256
\bar{H}_{i,-1,i,-1,i}+64 \bar{H}_{i,-1,i,-1,1}-360 \bar{H}_{i,-1,i,i,-1}+384
\bar{H}_{i,-1,i,i,i}-24 \bar{H}_{i,-1,i,i,1} \\ +64 \bar{H}_{i,-1,i,1,-1}-64
\bar{H}_{i,-1,i,1,i}-48 \bar{H}_{i,-1,1,-1,-1}+64 \bar{H}_{i,-1,1,-1,i}-16
\bar{H}_{i,-1,1,-1,1}+84 \bar{H}_{i,-1,1,i,-1} \\ -80 \bar{H}_{i,-1,1,i,i}-4
\bar{H}_{i,-1,1,i,1}-16 \bar{H}_{i,-1,1,1,-1}+16 \bar{H}_{i,-1,1,1,i}+192
\bar{H}_{i,i,-1,-1,-1}-256 \bar{H}_{i,i,-1,-1,i} \\ +64 \bar{H}_{i,i,-1,-1,1}-384
\bar{H}_{i,i,-1,i,-1}+368 \bar{H}_{i,i,-1,i,i}+16 \bar{H}_{i,i,-1,i,1}+64
\bar{H}_{i,i,-1,1,-1}-64 \bar{H}_{i,i,-1,1,i} \\ -280 \bar{H}_{i,i,i,-1,-1}+320
\bar{H}_{i,i,i,-1,i}-40 \bar{H}_{i,i,i,-1,1}+608 \bar{H}_{i,i,i,i,-1}-576
\bar{H}_{i,i,i,i,i}-32 \bar{H}_{i,i,i,i,1} \\ -40 \bar{H}_{i,i,i,1,-1}+32
\bar{H}_{i,i,i,1,i}+8 \bar{H}_{i,i,i,1,1}+64 \bar{H}_{i,i,1,-1,-1}-64
\bar{H}_{i,i,1,-1,i}-128 \bar{H}_{i,i,1,i,-1} \\ +112 \bar{H}_{i,i,1,i,i}+16
\bar{H}_{i,i,1,i,1}-48 \bar{H}_{i,1,-1,-1,-1}+64 \bar{H}_{i,1,-1,-1,i}-16
\bar{H}_{i,1,-1,-1,1}+84 \bar{H}_{i,1,-1,i,-1} \\ -80 \bar{H}_{i,1,-1,i,i}-4
\bar{H}_{i,1,-1,i,1}-16 \bar{H}_{i,1,-1,1,-1}+16 \bar{H}_{i,1,-1,1,i}+64
\bar{H}_{i,1,i,-1,-1}-64 \bar{H}_{i,1,i,-1,i} \\ -136 \bar{H}_{i,1,i,i,-1}+128
\bar{H}_{i,1,i,i,i}+8 \bar{H}_{i,1,i,i,1}-16 \bar{H}_{i,1,1,-1,-1}+16
\bar{H}_{i,1,1,-1,i}+20 \bar{H}_{i,1,1,i,-1} \\ -16 \bar{H}_{i,1,1,i,i}-4
\bar{H}_{i,1,1,i,1}+12 \bar{H}_{1,-1,-1,i,-1}-12 \bar{H}_{1,-1,-1,i,i}+6
\bar{H}_{1,-1,i,-1,-1}-16 \bar{H}_{1,-1,i,-1,i} \\ +10 \bar{H}_{1,-1,i,-1,1}-16
\bar{H}_{1,-1,i,i,-1}+16 \bar{H}_{1,-1,i,i,i}+10 \bar{H}_{1,-1,i,1,-1}-8
\bar{H}_{1,-1,i,1,i}-2 \bar{H}_{1,-1,i,1,1} \\ +12 \bar{H}_{1,-1,1,i,-1}-12
\bar{H}_{1,-1,1,i,i}-24 \bar{H}_{1,i,-1,-1,-1}+24 \bar{H}_{1,i,-1,-1,i}+36
\bar{H}_{1,i,-1,i,-1}-16 \bar{H}_{1,i,-1,i,i} \\ -20 \bar{H}_{1,i,-1,i,1}-8
\bar{H}_{1,i,-1,1,i}+8 \bar{H}_{1,i,-1,1,1}+32 \bar{H}_{1,i,i,-1,i}-32
\bar{H}_{1,i,i,-1,1}-24 \bar{H}_{1,i,i,i,-1} \\ +24 \bar{H}_{1,i,i,i,1}-32
\bar{H}_{1,i,i,1,-1}+32 \bar{H}_{1,i,i,1,i}-8 \bar{H}_{1,i,1,-1,i}+8
\bar{H}_{1,i,1,-1,1}-12 \bar{H}_{1,i,1,i,-1} \\ +16 \bar{H}_{1,i,1,i,i}-4
\bar{H}_{1,i,1,i,1}+8 \bar{H}_{1,i,1,1,-1}-8 \bar{H}_{1,i,1,1,i}+12
\bar{H}_{1,1,-1,i,-1}-12 \bar{H}_{1,1,-1,i,i} \\ +6 \bar{H}_{1,1,i,-1,-1}-16
\bar{H}_{1,1,i,-1,i}+10 \bar{H}_{1,1,i,-1,1}-16 \bar{H}_{1,1,i,i,-1}+16
\bar{H}_{1,1,i,i,i}+10 \bar{H}_{1,1,i,1,-1} \\ -8 \bar{H}_{1,1,i,1,i}-2
\bar{H}_{1,1,i,1,1}+12 \bar{H}_{1,1,1,i,-1}-12 \bar{H}_{1,1,1,i,i}\Big\}\, . \\
\end{multline}
The obtained results are in complete agreement with previous results at fixed spin values \cite{ABJMQSC12loops,Beccaria1,Beccaria2}. Note, that our $\barH$ - sums here  can be further rewritten using cyclotomic or S-sums of Ref. \cite{Ablinger3,Ablinger4} provided one extends the definition of the latter for the complex values of $x_i$ parameters. It is also possible to express them in terms of twisted $\eta$-functions introduced in  Ref. \cite{cuspQSC}. Here, we see that the maximal transcendentality principle\footnote{Similar considerations in the evaluation of Feynman diagrams first appeared in Ref. \cite{KotikovVeretin}. See also Ref. \cite{Tolya1,Tolya2}.} \cite{N4SYM2loop4,N4SYM3loop} also holds for anomalous dimensions of ABJM theory with the account for finite size corrections up to six loop order and it is now natural to assume it is validity for ABJM model to all orders. That is the results for anomalous dimensions in each order of perturbation theory are expressed in terms of $\barH$-sums of uniform weight $w$, where $w=3$ at NLO and $w=5$ at NNLO. In general,  the size of the basis of $\barH$-sums at weight $w$ is equal to $3\cdot 4^{w-1}$ and at NNNLO $(w=7)$ we should have 12288 such sums. Moreover, while discussing solution of NNLO constraints in subsection \ref{constraintsSolution} we noted that at present we are missing automatic reduction of $ \left\langle w_{1}\left(\j\right),w_{2}\left(\j\right),\ldots,w_{n}\left(\j\right)\right\rangle$ sums, arising at different steps of our calculation, to $\barH$ - sums, which makes intermediate expressions even larger. We are planing to address this latter issue in one of our subsequent publications. In addition it is desirable to construct Gribov-Lipatov reciprocity respecting basis \cite{GLreciprocity1,GLreciprocity2,GLreciprocity3} of generalized harmonic sums also for ABJM model. The latter in the case of $\mathcal{N}=4$ SYM is known to be much more compact compared  to  the original basis of harmonic sums and was used in Ref. \cite{Twist3SL2reciprocity,N4SYM5loop,N4SYM6looptwist3,N4SYM6loop,N4SYM7loop} to simplify the reconstruction of the full spin $S$ dependence of anomalous dimensions from the knowledge of anomalous dimensions at a set of fixed spin values.

\section{Conclusion}\label{Conclusion}

In this paper we have presented an algorithmic perturbative solution of ABJM quantum spectral curve for the case of twist 1 operators in $sl(2)$ sector of the theory. The solution treats operator spin $S$ as a symbol and applies to all orders of perturbation theory. The presented solution is performed directly in spectral parameter $u$-space and effectively reduces the solution of multiloop Baxter equations given by inhomogeneous second order difference equations with complex hypergeometric functions to purely algebraic problem. The solution is based on the introduction of a new class of functions - products of rational functions in spectral parameter with sums of Baxter polynomials and Hurwitz functions, which is closed under elementary operations, such as shifts and partial fractions,  as well as differentiation.  This class of functions is also sufficient for finding solutions of inhomogeneous Baxter equations involved. For the latter purpose we present recursive construction of $\FF_{\pm1}$ images for different products of Hurwitz functions with arbitrary indexes or fractions $\frac{1}{(u\pm i/2)^a}$ with leading order Baxter polynomials or their sums. The latter are entering inhomogeneous pieces of multiloop Baxter equations at different orders of perturbative expansion in coupling constant. Similar to Ref. \cite{VolinPerturbativeSolution,ABJMQSC12loops}, where all the operations performed were closing on trilinear combinations of rational, $\eta$ and $\P_k$ - functions,  all our operations are closing on fourlinear combinations of rational, $\eta$, $\P_k$ and $\langle Q|W\rangle$ - functions. As a particular application of our method we have considered anomalous dimensions of twist 1 operators in ABJM theory up to six loop order.  The obtained result was expressed in terms of generalized harmonic sums decorated by the fourth root of unity factors and introduced by us earlier. The results for anomalous dimensions respect the principle of maximum transcendentality.  It should be noted, that there is still a room for improvements of the proposed algorithm related to the simplifications of arising sums at different steps of presented solution. The advanced techniques for their reduction to $\barH$-sums will be the subject of one of our subsequent papers.

We expect the presented method to be generalizable to higher twists as well as to other theories, such as $\mathcal{N}=4$ SYM. 
The developed techniques should be also applicable for solution of twisted $\mathcal{N}=4$ and ABJM quantum spectral curves with $\bP$ functions having twisted non-polynomial asymptotic at large spectral parameter values, see \cite{twistedN4SYMQSC,QSCetadeformed,KonishiFate} and references therein. The latter models received recently a lot of attention in connection with the advances in so called fishnet theories  \cite{fishnet1,Isaev,fishnet2,fishnet3,fishnet4,fishnet5,fishnet6,fishnet7,fishnet8,fishnet9,fishnet10,fishnet11,fishnet12,fishnet13,fishnet14,fishnet15}. Moreover, similar ideas should be also applicable to the study of BFKL regime within quantum spectral curve approach \cite{GromovBFKL1,GromovBFKL2,BFKLnonzeroConformalSpin} for $\mathcal{N}=4$ SYM. In the latter case we also have a perturbative expansion when both coupling constant $g$ and parameter $w\equiv S+1$, describing the proximity of operator spin $S$ to $-1$ are considered to be small, while  their ratio $g^2/w$ remains fixed.

\section*{Acknowledgements}

This work was supported by Foundation for the Advancement of Theoretical Physics and Mathematics "BASIS". 

\appendix

\section{Hurwitz functions}\label{Hurwitz-functions}

 We define Hurwitz functions entering the presented solution as 
 \begin{align}\label{eq:xiseries}
 \xi_{\left|a\right|,A} & =\sum_{n=1}^{\infty}\frac{1}{\left(u+in-\frac{i}{2}\right)^{\left|a\right|}}\xi_{A}^{\left[2n\right]}\\
 \xi_{-\left|a\right|,A} & =-\sum_{n=1}^{\infty}\frac{\left(-\right)^{n}}{\left(u+in-\frac{i}{2}\right)^{\left|a\right|}}\xi_{A}^{\left[2n\right]} ,
 \end{align}
Here $A$ denotes the arbitrary sequence of indexes and $\xi$ function without indexes is identical to unity.  These are the shifted versions of Hurwitz functions introduced in \cite{VolinPerturbativeSolution,ABJMQSC12loops}
 \begin{equation}
 \xi_{A}=\eta_{A}^{\left[1\right]}
 \end{equation}
 The $\xi_{1\ldots 1}$ functions should be defined separately, as the series \eqref{eq:xiseries} diverge in this case. For $\xi_1$ - function we have 
 \begin{equation}
 \xi_{1}\left(u\right)=i\psi\left(-iu+\tfrac{i}{2}\right)\,.
 \end{equation}
and $\xi_{1\ldots 1}$ functions are defined as \cite{VolinMZVdoublewrapping}:
\begin{equation}
\xi_{\underbrace{1,\ldots,1}_{k}}\left(u\right)=\frac{1}{k!}\left(\xi_{1}+\partial_{u}\right)^{k}1\,.
\end{equation} 	
For shifts of our Hurwitz functions we have
\begin{align}\label{eq:xishift}
\xi_{a,A}^{\left[2\right]} & =\sigma_{a}\xi_{a,A}-\frac{\sigma_{a}}{\left(u+\frac{i}{2}\right)^{\left|a\right|}}\xi_{A}^{\left[2\right]}\\
\xi_{a,A}^{\left[-2\right]} & =\sigma_{a}\xi_{a,A}+\frac{1}{\left(u-\frac{i}{2}\right)^{\left|a\right|}}\xi_{A}.
\end{align}

\section{H and B - sums}\label{H-B-sums}

To write down the results for anomalous dimensions we introduce generalization of harmonic sums decorated with the fourth root of unity factors $(\exp(i\pi/2))^n$ (indexes may be either real or purely imaginary) 
\begin{equation}
H_{a,b,\ldots}(S) = \sum_{k=1}^S \frac{\Re [(a/|a|)^k]}{k^{|a|}} H_{b,\ldots} (k)\, \quad H_{a,\ldots} = H_{a,\ldots} (S)\, \quad \barH_{a,\ldots} = H_{a,\ldots} (2S)\,
\end{equation}
In addition our intermediate expressions for coefficients in the anzats for $\bP$ functions and for constants entering homogeneous pieces  of solutions for $\nu_{1,2}^{(i)}$ functions contain $B$ - sums. The latter are defined similar to\footnote{See appendix B there.}  \cite{ABJM_QSC_Mellin}. That is we have  
 $B_0 = B_0(S)=1$, $B_1 = B_1(S)=H_1(S)-H_{-1}(S)$, and $B_{n>1}$ is defined recursively by the symbolic formula
 \begin{equation}
 B_{n}=\left(O_{1}+\left(-1\right)^{n}O_{-1}\right)B_{n-1}\,,
 \end{equation}
 where $O_{\pm1}$ is a linear operator prepending index $\pm 1$ to harmonic sums, i.e. $O_{\pm1}H_{\boldsymbol{a}}\left(S\right)=H_{\pm1,\boldsymbol{a}}\left(S\right)$.
 In particular, we have
 \begin{align}
 B_{2}=&\left(O_{1}+O_{-1}\right)B_{1}=H_{1,1}+H_{-1,1}-H_{1,-1}-H_{-1,-1}\,,\\
 B_{3}=&\left(O_{1}-O_{-1}\right)B_{2}=H_{1,1,1}+H_{1,-1,1}-H_{1,1,-1}-H_{1,-1,-1}\\
 &-H_{-1,1,1}-H_{-1,-1,1}+H_{-1,1,-1}+H_{-1,-1,-1}\,.
 \end{align}

\bibliographystyle{hieeetr}
\bibliography{litr}

\end{document}